\documentclass[aps,twocolumn,floats,superscriptaddress,prd,nofootinbib]{revtex4-1}
\usepackage{graphicx, epsfig, bm, amsmath}
\def\mnras{Mon.~Not.~Roy.~Astron.~Soc.}
\usepackage{color}
\usepackage{hyperref}

\begin{document}


\newcommand{\mmin}{M_{\rm min}}
\newcommand{\hmsol}{h^{-1}\,M_{\odot}}
\newcommand{\msol}{M_{\odot}}
\newcommand{\nmz}{\bar N(M,z)}
\newcommand{\fsky}{f_{\rm sky}}

\newcommand{\zmin}{z_{\rm min}}
\newcommand{\zmax}{z_{\rm max}}
\newcommand{\dz}{\Delta z}
\newcommand{\dzsub}{\Delta z_{\rm sub}}
\newcommand{\zminsn}{z_{\rm min}^{\rm SN}}
\newcommand{\nzpc}{N_{z,{\rm PC}}}
\newcommand{\nmax}{N_{\rm max}}
\newcommand{\amax}{a_{\rm max}}
\newcommand{\atr}{a_{\rm tr}}
\newcommand{\aeq}{a_{\rm eq}}

\newcommand{\wmin}{w_{\rm min}}
\newcommand{\wmax}{w_{\rm max}}
\newcommand{\wfid}{w_{\rm fid}}

\newcommand{\lcdm}{$\Lambda$CDM}

\newcommand{\om}{\Omega_{\rm m}}
\newcommand{\ode}{\Omega_{\rm DE}}
\newcommand{\ok}{\Omega_{\rm K}}
\newcommand{\omhh}{\Omega_{\rm m} h^2}
\newcommand{\obhh}{\Omega_{\rm b} h^2}
\newcommand{\winf}{w_{\infty}}
\newcommand{\scrm}{\mathcal{M}}
\newcommand{\osf}{\Omega_{\rm sf}}
\newcommand{\omf}{\Omega_{\rm m}^{\rm fid}}
\newcommand{\scrmf}{\mathcal{M}^{\rm fid}}
\newcommand{\rhode}{\rho_{\rm DE}}
\newcommand{\rhoc}{\rho_{{\rm cr},0}}
\newcommand{\rhom}{\rho_{{\rm m},0}}
\newcommand{\dlnl}{-2\Delta\ln\mathcal{L}}
\newcommand{\dlum}{d_{\rm L}}
\newcommand{\dlss}{D_*}
\newcommand{\zh}{z_h}
\newcommand{\zbao}{z_{\rm BAO}}
\newcommand{\gpr}{G^{\prime}}

\newcommand{\fnl}{f_{\rm NL}}

\definecolor{darkgreen}{cmyk}{0.85,0.2,1.00,0.2}
\newcommand{\mm}[1]{\textcolor{red}{[{\bf MM}: #1]}}
\newcommand{\dhcomment}[1]{\textcolor{darkgreen}{[{\bf DH}: #1]}}
\newcommand{\wh}[1]{\textcolor{blue}{[{\bf WH}: #1]}}

\newcommand{\thetal}{\bm{\theta}_{\Lambda}}
\newcommand{\thetaq}{\bm{\theta}_{\rm Q}}
\newcommand{\thetade}{\bm{\theta}_{\rm DE}}
\newcommand{\thetas}{\bm{\theta}_{\rm S}}
\newcommand{\thetadel}{\bm{\theta}_{{\rm DE},\Lambda}}
\newcommand{\thetadeq}{\bm{\theta}_{\rm DE,Q}}
\newcommand{\thetades}{\bm{\theta}_{\rm DE,S}}
\newcommand{\thetaother}{\bm{\theta}_{\rm nuis}}
\newcommand{\Mobs}{M_{\rm obs}}
\newcommand{\aap}{Astron. Astrophys.}


\pagestyle{plain}

\title{Simultaneous falsification of $\Lambda$CDM and quintessence with massive, distant clusters}

\author{Michael J.\ Mortonson}
\affiliation{Center for Cosmology and AstroParticle Physics, 
        The Ohio State University, Columbus, OH 43210}

\author{Wayne Hu}
\affiliation{Kavli Institute for Cosmological Physics, Department of Astronomy \& Astrophysics,
       and Enrico Fermi Institute,
        University of Chicago, Chicago, IL 60637}

\author{Dragan Huterer}
\affiliation{Department of Physics, University of Michigan, 
450 Church St, Ann Arbor, MI 48109-1040}

\begin{abstract}
Observation of even a single massive cluster, especially at high redshift,
can falsify the standard cosmological
framework consisting of a cosmological constant and cold dark matter (\lcdm)
with Gaussian initial conditions by exposing an inconsistency between the
well-measured expansion history and the growth of structure it predicts.
Through a likelihood analysis of 
current cosmological data that constrain the expansion history, we show that
the \lcdm\ upper limits on the expected number of massive, distant clusters
are nearly identical to limits predicted by {\it all} quintessence models
where dark energy is a minimally coupled scalar field with a canonical
kinetic term.  
We provide convenient fitting formulas for the confidence level at which
the observation of a cluster of mass $M$ at redshift $z$ can falsify \lcdm\
and quintessence given cosmological parameter uncertainties and sample
variance, as well as for the expected number of such clusters in the light cone and the Eddington bias factor that must be 
applied to observed masses.  By our conservative confidence criteria, which
equivalently require masses $3$ times larger than typically
expected in surveys of a few hundred square degrees, none of the presently
known clusters falsify these models.
Various systematic errors, including uncertainties 
in the form of the mass function and differences between supernova light curve fitters, 
typically shift the exclusion curves by less than $10\%$ in mass, making current statistical and systematic uncertainties in cluster mass determination the most critical factor in assessing falsification of \lcdm\ and quintessence. 
\end{abstract}

\maketitle

\section{Introduction}
\label{sec:intro}

It is well known that the presence of even a single high redshift cluster with
sufficient mass can falsify the standard cosmological model where such objects
grow from Gaussian initial conditions under the gravitational instability of
cold dark matter in a cosmological constant dominated universe (\lcdm)
\cite{Bahcall:1998ur,Haiman:2000bw,Weller:2001gk,Oguri:2008ww,Holz:2010ck,OukbirBlanchard}.
Robust upper bounds on the number of high mass clusters in \lcdm\ arise from
the exponential suppression of the dark matter halo number density with mass
and the fact that, in the \lcdm\ paradigm, geometric constraints on the
expansion history determine the growth of structure.  Indeed, recently detected massive
clusters at high redshift \cite{Mullis:2005hp,Lombardi:2005nd,Brodwin:2010ig}
have led to claims of 
tension with
\lcdm\ \cite{Jee:2009nr,Holz:2010ck,Cayon,Hoyle}.

More generally, for any given paradigm for dark energy, geometric constraints in combination with CMB constraints on the initial amplitude of fluctuations
can be translated into upper bounds on the abundance of high mass clusters.
Observational violation of these upper bounds would therefore falsify {\it
  all} models of that given paradigm.  In particular, in previous work
\cite{Mortonson:2009hk} we established that the linear growth function of all
quintessence models, where dark energy is a canonical, minimally coupled
scalar field, is bounded above to be within a few percent of the
\lcdm\ values.  Here we translate these upper bounds on the linear growth rate
to upper bounds on the number of clusters above a given mass and redshift.  In
particular, we make a conservative assessment of the limiting mass and
redshift of a cluster that would rule out all quintessence models that is
robust to our present knowledge of cosmological parameters, supernova light
curve fitters, sample variance, and simulation-based calibration of the
cluster abundance.

The predictions we present here incorporate cosmological constraints 
from several recent data sets. In addition to placing limits on  
the expansion history of the universe, these data also provide 
important information about the amplitude of matter density fluctuations which 
directly feeds into cluster predictions.
In particular, observations of 
cosmic microwave background (CMB) anisotropy constrain the amplitude 
of perturbations at the epoch of last scattering, $z=1090$.
In contrast to many previous studies of the effects of dark energy on 
the growth of structure,
we do \emph{not} take this constraint on the 
density fluctuations at early times to mean that the amplitude of 
perturbations at $z=0$, often characterized by the parameter $\sigma_8$, 
is also well constrained. Instead, we combine CMB constraints with 
the linear growth functions of quintessence models to set $\sigma_8$ so 
that these models are fully consistent with present CMB data.

We begin in \S \ref{sec:methods} by developing methodology to extract
cluster abundance probability distributions
from expansion history measurements in the context of
a given dark energy paradigm.   In \S \ref{sec:predictions} we show how to
convert these distributions into confidence levels at which a cluster of a given mass
and redshift excludes \lcdm\ and quintessence.  We consider 
statistical errors due to parameter
and sample variance as well as systematic shifts from uncertainties in 
observational mass determination, the form of the mass function, 
and the analysis of supernova data.
We discuss these results in \S \ref{sec:discussion}.  

In Appendix \ref{sec:fit} we provide convenient fitting functions for the
relationships between cluster masses, redshifts, numbers, and confidence levels, as well as
the slope of the mass function for bias corrections. 
In Appendix \ref{sec:ede} we discuss the impact of the CMB normalization of structure, especially in the context of early dark energy.   
Finally, in Appendix \ref{sec:eddingtonbias} we discuss two different types of biases induced by measuring observable proxies for mass in the presence of
a steep mass function.

\section{Methodology}
\label{sec:methods}

Following Refs.~\cite{Mortonson:2008qy,Mortonson:2009hk}, we use current constraints
on the expansion history of the Universe to make falsifiable predictions for
observables related to the growth of structure under specific dark energy
paradigms.  Here we briefly summarize this technique and highlight changes
in the methodology that enable us to obtain robust predictions for the 
cluster abundance. We begin with descriptions of the data sets we use to 
constrain the expansion history and the initial amplitude of 
density fluctuations (\S \ref{sec:data}), followed by a summary of the likelihood analysis 
that determines which models in the \lcdm\ and quintessence 
paradigms satisfy the observational constraints (\S \ref{sec:MCMC}). Finally,
we show how we use the output of this analysis to compute probabilities 
for the abundance of massive, distant clusters in the context of 
various dark energy paradigms (\S \ref{sec:abundance}).

\subsection{Data Sets}
\label{sec:data}

The main observational constraints that inform our predictions for
cluster abundances are relative distance measures from Type Ia 
supernovae (SNe), the
CMB temperature and polarization power spectra, baryon acoustic oscillation
(BAO) distance measures, and local distance measures of the Hubble constant 
($H_0$).

The Type Ia SN sample we use is the compilation of 288 SNe from
Ref.~\cite{SDSS_SN}, consisting of data from the first season of the Sloan
Digital Sky Survey-II (SDSS-II) Supernova Survey, the ESSENCE survey
\cite{WoodVasey_2007}, the Supernova Legacy Survey \cite{Astier}, Hubble Space
Telescope SN observations \cite{Riess_2006}, and a collection of nearby SN
data \cite{Jha:2006fm}.  The light curves of these SNe have been uniformly
analyzed by \cite{SDSS_SN} using both the MLCS2k2 \cite{Jha:2006fm} 
and SALT2 \cite{Guy:2007dv} methods.
We use the MLCS2k2-analyzed data for most of our results since
it leads to the more conservative bound on massive clusters. 
For example, for flat \lcdm\, using MLCS2k2 SN data
increases $\sigma_8 \Omega_{\rm m}^{0.5}$ by $\sim 7\%$ relative to SALT2.  
In \S~\ref{sec:syst} we address the impact of the choice of SN 
analysis method and also compare with constraints 
from the Union2 compilation \cite{Amanullah:2010vv} of 557 SNe, 
which includes the CfA3 sample \cite{Constitution} and a number of SN data sets 
previously combined in the first Union compilation \cite{SCP_Union}.

For the CMB, we use the most recent, 7-year release of data from the WMAP
satellite (WMAP7) \cite{Larson:2010gs} employing a modified version of the likelihood
code available at the Legacy Archive for Microwave Background Data Analysis Web site \cite{WMAP_like} which is substantially
faster than the standard version while remaining sufficiently 
accurate \cite{Dvorkin:2010dn,wmapfast_url}.  We
compute the CMB angular power spectra using the code CAMB
\cite{Lewis:1999bs,camb_url} modified with the parametrized post-Friedmann
(PPF) dark energy module \cite{PPF,ppf_url} to include models with general
dark energy equation of state evolution where $w(z)$ may cross $w=-1$.

We use the BAO constraints from Ref.~\cite{PercivalBAO}, which combines 
data from SDSS and the 2-degree Field Galaxy Redshift Survey that
determine the ratio of the sound horizon at last scattering to the 
quantity $D_V(z)\equiv [zD^2(z)/H(z)]^{1/3}$ at redshifts $z=0.2$ and $z=0.35$.
Since these constraints actually come from galaxies spread over a range 
of redshifts, and our most general dark energy model classes allow the 
possibility of significant variations in $H(z)$ and $D(z)$ across this
range, we implement the constraints by taking the volume average of $D_V$ 
over $0.1<z<0.26$ (for $z=0.2$) and $0.2<z<0.45$ (for $z=0.35$).
The effect of this volume averaging on the final combined constraints 
from current data is relatively small.

Finally, we include the recent Hubble constant measurement from the SHOES team
\cite{SHOES}, based on SN distances at $0.023<z<0.1$ that are linked to a
maser-determined absolute distance using Cepheids observed in both the maser
galaxy and nearby galaxies hosting Type Ia SNe.  The SHOES measurement
determines the absolute distance to a mean SN redshift of $z=0.04$
which we implement as 
$D(z=0.04) = 0.04c/(74.2 \pm 3.6$~km~s$^{-1}$~Mpc$^{-1}$).

\subsection{MCMC Analysis}
\label{sec:MCMC}

To predict the cluster abundance using constraints from current
data, we use a Markov Chain Monte Carlo (MCMC) likelihood analysis.  We take a
set of parameters $\bm{\theta}$ that completely describes a given dark energy class
and use a modified version of the code CosmoMC \cite{Lewis:2002ah,cosmomc_url}
to sample from the joint posterior distribution of the parameters,
\begin{equation}
{\cal P}(\bm{\theta}|{\bf x})=
\frac{{\cal L}({\bf x}|\bm{\theta}){\cal P}(\bm{\theta})}{\int d\bm{\theta}~
{\cal L}({\bf x}|\bm{\theta}){\cal P}(\bm{\theta})},
\label{eq:bayes}
\end{equation}
where ${\cal L}({\bf x}|\bm{\theta})$ is the likelihood of the data ${\bf x}$
given the model parameters $\bm{\theta}$ and ${\cal P}(\bm{\theta})$ is the
prior probability density.
We test convergence of the samples to a stationary
distribution by applying a conservative Gelman-Rubin criterion
\cite{gelman/rubin} of $R-1\lesssim 0.01$ across a minimum of four chains for
each model class.

For the \lcdm\ class we take the parameters
\begin{eqnarray}
\thetal &=&\{\om, \ok, \omhh, \obhh, n_s, \ln A_s, \tau \} \,.
\label{eq:lcdmpar}
\end{eqnarray}
We will mainly consider the flat \lcdm\ class here where $\ok=0$.
Additional parameters such as $H_0$ and $\ode$ are derived from this 
fundamental set.
In particular, the present amplitude of the linear power spectrum
$\sigma_8$ is a derived parameter (see Appendix \ref{sec:ede}).
Our normalization parameter is $A_s$, the amplitude of the {\em initial} 
curvature power spectrum at $k=0.05$\,Mpc$^{-1}$.  
For all parameters in Eq.~(\ref{eq:lcdmpar}) we take flat priors
that are wide enough that they do not limit the MCMC constraints from 
current data.

For quintessence we extend the parameter set of Eq.~(\ref{eq:lcdmpar}) by
taking a principal component (PC) decomposition of the
dark energy equation of state for $z<1.7$,
\begin{equation}
w(z) +1  = \sum_{i=1}^{\nmax} \alpha_i e_i(z) \, ,
\label{eq:pcstow}
\end{equation}
where $\alpha_i$ are the PC amplitudes, $\nmax = 10$ is the number of
components required to form a complete basis with respect to growth and
distance measures \cite{Mortonson:2009hk}.  These principal components are
constructed from the eigenvectors of a projection for the Planck CMB and 
SuperNova Acceleration Probe SN covariance matrix for $w(z)$ in sufficiently 
fine redshift bins to approximate continuous equation of state variations, 
as described in detail in Ref.~\cite{Mortonson:2008qy}.

We parametrize the dark energy equation of state at $z>1.7$ by 
a constant,
\begin{equation}
w(z > 1.7) = \winf \,.
\end{equation}
While this parametrization does not completely describe all possible
behaviors for the equation of state, it does allow for dark energy that is a
non-negligible fraction of the total at high redshift or ``early dark
  energy'' (EDE).  For more restricted model classes without EDE, we fix
$\winf=-1$ since a constant dark energy density rapidly becomes negligible
relative to the matter density at increasing redshift.

In summary, our quintessence parameters are
\begin{eqnarray}
\thetaq &=& \{ \thetal, \alpha_1,\ldots, \alpha_{10}, \winf\} \,.
\label{eq:parametersfull}
\end{eqnarray}
Note that flat \lcdm\ is a special case of quintessence with
$\{\alpha_i,1+w_\infty,\ok\}=0$. We also consider a restricted
quintessence class of models which are flat and do not have significant EDE,
corresponding to $\{1+w_\infty,\ok\}=0$.  Quintessence models describe
dark energy as a scalar field with kinetic and potential contributions to
energy and pressure.  Barring models where large kinetic and (negative)
potential contributions cancel (e.g.~\cite{Mortonson:2009qq}), quintessence equations of state are restricted
to $-1\leq w(z)\leq 1$.  Following \cite{Mortonson:2008qy}, this bound is conservatively
implemented with independent top-hat priors on the PC amplitudes $\alpha_i$.
Any combination of PC amplitudes that is rejected by these priors must arise
from an equation of state that violates the bound on $w(z)$, but not
all models that are allowed by the priors strictly satisfy this bound.  This
prior is thus appropriate for making conservative statements on the
falsifiability of quintessence.  For EDE, quintessence requires $\winf \ge
-1$, and we additionally impose $\winf \le 0$ to maintain the usual matter and
radiation dominated epochs at high redshift. We adopt a flat prior on 
$\exp(\winf)$ which gives greater weight to models with $\winf$ near 0.

\subsection{Cluster Abundance}
\label{sec:abundance}

As described in \cite{Mortonson:2008qy}, the MCMC approach allows us to
straightforwardly calculate confidence regions for observable quantities
determined by the evolution of large-scale structure. The first step is to
compute the posterior probability of the linear growth function $G(z)$ from the
joint posterior of the dark energy parameters. Note that the growth function
we use here scales out the growth of density perturbations during matter
domination, $\delta\propto a$, so $G(z)\propto (1+z)\delta$ with normalization
$G(z=10^3)=1$.

Given the predicted growth function, we compute the abundance of clusters by integrating the product of the
halo mass function $dn/d\ln M$ and the comoving volume element over cluster 
mass and redshift. Since we are interested in the most massive and 
most distant clusters, we integrate above thresholds in mass and redshift 
to obtain the expected number of clusters in the full sky 
with mass $>M$ and redshift $>z$,
\begin{equation}
\bar N(M,z) =  \int_{z}^{\infty} dz' \frac{4\pi D^2(z')}{H(z')}
\int_{M}^{\infty} \frac{dM'}{M'} \frac{dn}{d\ln M}(M',z')\,,
\label{eq:navg}
\end{equation}
where $4\pi D^2/H$ is the comoving volume element for the full sky written in
terms of the comoving angular diameter distance $D(z)$ and the Hubble
expansion rate $H(z)$. Note that since the high mass, high redshift 
mass function falls off rapidly with increasing mass and redshift, 
$\bar N(M,z)$ is typically 
dominated by the abundance near the threshold values of $M$ and $z$.

The form of the mass function can be inferred by 
fitting to the abundance of dark matter halos identified in numerical simulations.
The cosmological dependence of the mass function is typically 
expressed as
\begin{equation}
\frac{dn}{d\ln M} = \frac{\rhom}{M} 
\left|\frac{d\ln\sigma}{d\ln M}\right| f(\sigma,z)\,,
\label{eq:mf1}
\end{equation}
where $\sigma(M,z)$ is
the rms of linear density fluctuations smoothed over spheres of comoving 
radius $R=(3 M/4\pi \rhom)^{1/3}$ 
and $\rhom=3\om H_0^2/(8\pi G)$ is the present matter density.
Here $f(\sigma,z)$ is a function determined by the fit to simulations 
that depends primarily on $\sigma(M,z)$, but is also
weakly dependent on redshift \cite{Tinker:2008ff,Bhattacharya:2010wy}.

We study the dependence of our predictions on the specific choice 
of mass function and describe those tests in the next section, 
but for definiteness we adopt the Tinker et al.\ \cite{Tinker:2008ff} 
mass function for our main results, 
\begin{equation}
f(\sigma,z) = A \left[ \left( {\sigma \over b}\right)^{-a} + 1
\right] e^{-c/\sigma^2},
\label{eq:mf2}
\end{equation}
with $A=0.186(1+z)^{-0.14}$, $a=1.47(1+z)^{-0.06}$, $b=2.57(1+z)^{-0.011}$,
$c=1.19$.  This fit assumes $M=M_{200}$, defined as the mass within a
spherical region around the halo center enclosing an average density equal to
200 times the mean matter density, $\rhom(1+z)^3$.  We do not adjust the
fit parameters for variations in the dark energy model but do test the
sensitivity to the fidelity of the fit in \S \ref{sec:syst}.

For the dark energy models and range of scales and redshift that we consider here, 
the linear growth function  is approximately scale-independent, so 
we can separate the mass and redshift dependence of the density 
fluctuation rms as
\begin{equation}
\sigma(M,z) = \sigma(M,0) \frac{G(z)}{(1+z)G(0)}\,.
\end{equation}
We obtain $\sigma(M,0)$ for each cosmological model using the modified 
version of CAMB, and compute $G(z)$ by integrating the differential 
equation for the linear growth function as described in \cite{Mortonson:2008qy}.
By using the growth function only to scale backwards from the present epoch,
this method includes all high redshift modifications to the transfer function and 
the CMB normalization of $\sigma(M,0)$ through $A_s$.  We only assume that
at the low redshifts of interest $0<z<2$ the growth function is independent of scale
(see Appendix \ref{sec:ede} for discussion of EDE clustering at early times).

\section{Cluster Predictions}
\label{sec:predictions}

From the complete parametrization of the \lcdm\ and quintessence
model classes and the MCMC posterior probability of the mean number of 
clusters $\bar N$ across the full sky above a given mass and redshift, we can assess the confidence with which the observation of a cluster with that mass and redshift 
can falsify the model class. We first consider the impact of parameter and 
statistical uncertainties on the predicted number of massive, high redshift 
clusters in \S\ref{sec:predict}.  In \S \ref{sec:exclusion} we combine these into
exclusion curves in mass and redshift and evaluate the significance of 
the most massive clusters in the present high redshift sample.   
Finally, in \S \ref{sec:syst} we illustrate how various systematic errors would shift 
the upper limits on massive cluster abundances.

\subsection{Parameter and Sample Confidence}
\label{sec:predict}

We quantify two types of confidence limits for statistical uncertainties.  The
first is associated with parameter uncertainties on the mean number $\bar N$
within the dark energy model class.  
We call this {\em{parameter variance}} and take the
one-tailed $100p\%$ confidence level (CL) upper limits on the mean number, 
$\bar N_{Pp}(M,z)$; for example, given the
CMB, SN, BAO, and $H_0$ constraints, there is a 95\% probability that the mean
number of clusters $>M$ and $>z$ in the full sky is less than $\bar N_{P.95}$.
To a good approximation, the parameter variance in our cluster abundance 
predictions corresponds to variance in $\sigma(M,z)$ or, in particular, 
$\sigma_8$. For example, flat \lcdm\ models at the 95\% parameter CL
have a larger amplitude of fluctuations, $\sigma_8\approx 0.87$, 
than the median models with $\sigma_8\approx 0.83$.

The second confidence limit we define is associated with {\em sample variance},
under the assumption that the number of
clusters in the sample is Poisson distributed with mean $\bar N$ across the
full sky.  This assumption ignores the clustering of clusters and should be a
good approximation in the rare object limit \cite{Hu:2002we}.  In particular,
the probability to have zero clusters in a random sample of a fraction of sky
$\fsky$ is $s \equiv e^{-\bar N \fsky}$.
We therefore define the sample variance
$100s\%$ CL for models with a mean number of clusters in the
\emph{full sky} as 
\begin{equation}
\bar N_{Ss}(\fsky) \equiv -\fsky^{-1} \ln s\,.
\end{equation}
That is, if the mean number of clusters above $M$ and $z$ 
expected in the full sky is 
$\bar N_{Ss}$ for a particular model, then that model would be 
excluded at the $100s\%$ CL by one or more such observed clusters in 
a survey covering $\fsky$ of the full sky.
For example, the observation of one or
  more clusters at redshift $z$ with mass $M$ observed in $300
\deg^2$ ($\fsky\approx 0.0073$) would exclude models that predict a mean number
of clusters in the full sky $\bar N(M,z)$ less than $\bar N_{S.95}\approx
7.1$ at the 95\% sample CL. 

Some care must be taken to define the appropriate $\fsky$ for a given cluster.
For example, if out of many similar surveys only one reported a high mass 
cluster, then the appropriate sky area
is the total area of the surveys, not just the individual survey area selected to 
have the cluster {\it a posteriori}.   The most conservative limits are obtained 
by taking $\fsky=1$ when interpreting
any observation, i.e. assuming that all unobserved regions of the sky do 
not host clusters with anomalously high masses and redshifts.  
In this case $\bar N_{S.95} \approx 0.051$.  Compared with,
say, the median prediction $\bar N_{S.50} = 95$ at 300 deg$^{2}$, these
criteria are a factor of $\sim 1900$ more conservative in predicted number.

We combine these two types of uncertainties to compute the
maximum cluster mass and redshift within some area of the sky predicted by a
particular model class. The mass and redshift limits corresponding to $100s\%$
sample CL and $100p\%$ parameter CL can be
found by taking $\bar N_{Ss}(\fsky)=\bar N_{Pp}(M,z)$ to get
\begin{equation}
\int_{-\infty}^{\log(-\fsky^{-1}\ln s)} d\log \bar N \,P(\log\bar N | M,z) = p\,,
\label{eq:cldef}
\end{equation}
where $P(\log\bar N | M,z)$ is the posterior density in the expected
  number of clusters above $M$ and $z$ for the given class of dark
  energy models.  For simplicity, we will often consider the case $s=p$ and
refer to this as the ``$100s\%$ joint CL'' for sample and parameter variance.
Note that in this approach observational uncertainty in determining
the mass, which varies from cluster to cluster, is not directly included so that
the cluster mass errors must be included when comparing with the $M(z)$ 
exclusion curves presented in \S \ref{sec:exclusion} 
(see also Appendix~\ref{sec:eddingtonbias}).

Finally, one can also  test  whether the
$N$ rarest clusters together place a substantially stronger bound on the
dark energy paradigm than the single rarest object detected so far.
 To do this, one computes the
Poisson probability using the mass and redshift threshold that includes 
all of those $N$
clusters.  For example, for the two rarest clusters the mean
number $\bar N_{Ss}^{(2)}$ corresponding to exclusion at the $100s\%$ sample CL
can be found from
\begin{eqnarray}
 s &=& \left(1+\fsky \bar N_{Ss}^{(2)}\right) e^{-\fsky \bar N_{Ss}^{(2)} } 
 \label{eq:samp2}
\end{eqnarray}
to be
\begin{eqnarray}
\bar N_{Ss}^{(2)} &=& -\fsky^{-1} [1+W_{-1}(s)]   \,,
\end{eqnarray}
where $W_{-1}$ is the lower branch of the Lambert $W$ function. If $\fsky \bar
N_{Ss}^{(2)} \ll 1$, $\bar N_{Ss}^{(2)} \approx \fsky^{-1}\sqrt{2(1-s)}$.  The
expected number for the 95\% joint CL at $\fsky=1$ is $\bar N_{S.95}^{(2)} =
0.355$, compared with $\bar N_{S.95}=0.051$ for the single most extreme
cluster; therefore, the model is required to predict a mean abundance 7 times 
larger to explain two clusters above a given mass and redshift rather than one.
This statistic is conservative in the sense that the rarest cluster is typically treated as
if it were only as rare as the second rarest cluster (more specifically both
clusters are assigned the lowest $M,z$ of the pair).  
Thus the $N=2$ test can actually be weaker than the $N=1$ rarest cluster test.  
It can be applied
sequentially to the $N$ rarest clusters by defining $\bar N_{Ss}^{(N)}$ as the
solution to
\begin{equation}
s = \sum_{i=0}^{N-1} { (\fsky \bar N_{Ss}^{(N)})^i \over i! } e^{-\fsky\bar N_{Ss}^{(N)}}.
\end{equation}
Again if  $\fsky \bar
N_{Ss}^{(N)} \ll 1$, $\bar N_{Ss}^{(N)} \approx \fsky^{-1}[N!(1-s)]^{1/N}$.
In the confidence level fitting formula of Eq.~(\ref{eq:logn}) one simply replaces
\begin{equation}
-\fsky^{-1}\ln s =  \bar N_{Ss} 
\rightarrow \bar N_{Ss}^{(N)}.
\end{equation}
Note however that the number of trials taken before finding an anomaly must be
considered in interpreting the exclusion.

\begin{figure}[t]
\centerline{\psfig{file=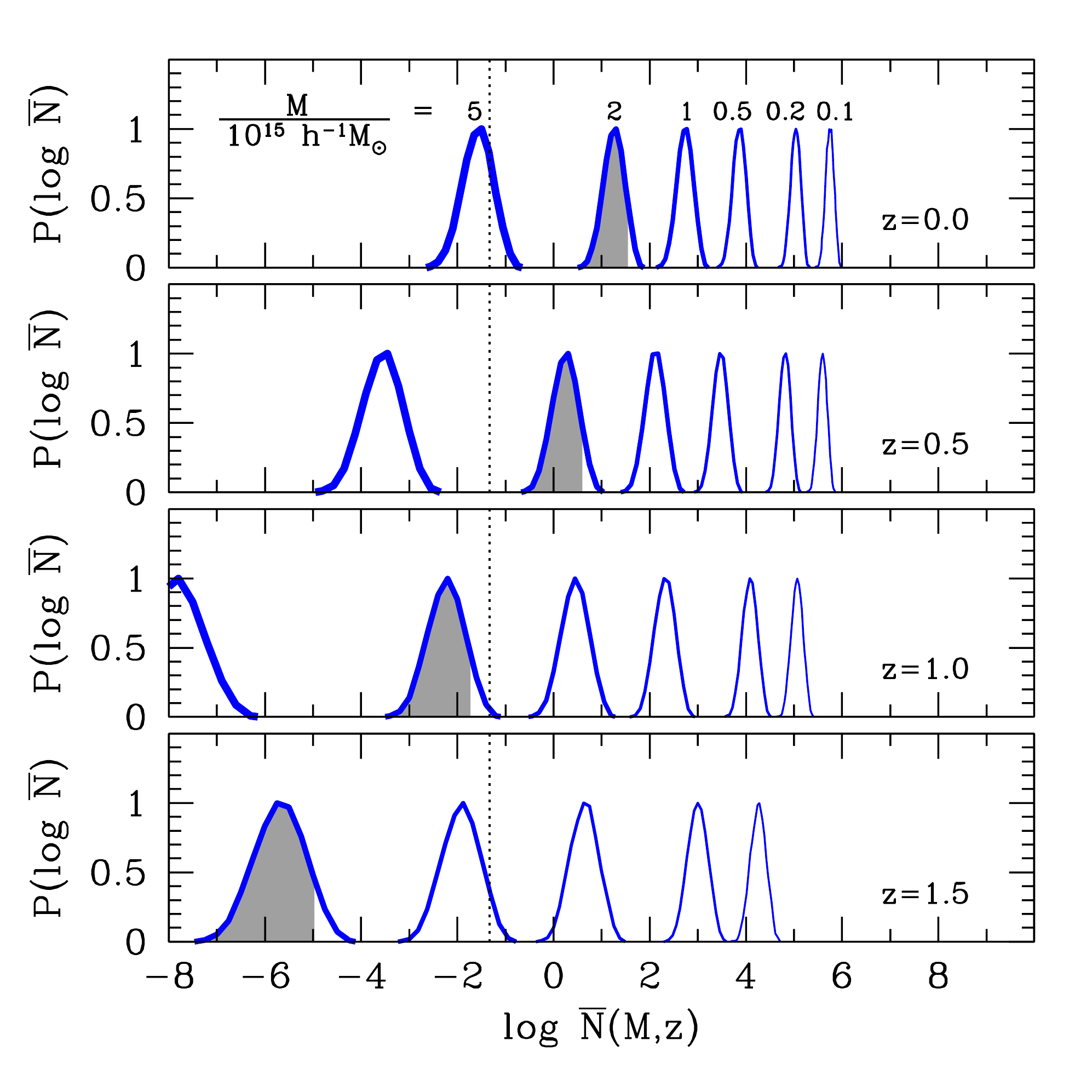, width=3.2in}}
\caption{\footnotesize Predicted mean, full-sky abundance $\nmz$ of clusters above mass and
  redshift thresholds $M$ and $z$, respectively, for flat \lcdm\ models that
  fit current CMB+SN+BAO+$H_0$ data. Vertical dotted lines are plotted 
  at $\bar N_{S.95}(\fsky=1)$, the 95\% CL sample variance limit for a
  full-sky survey.  For $M=2\times 10^{15}\,\hmsol$, 
  we shade the lower 95\% of each distribution; exclusion at the 
  95\% \emph{joint} CL for a cluster of this mass in the full sky occurs 
  at the redshift for which all of the shaded area lies to the left of the 
  vertical $\bar N_{S.95}(\fsky=1)$ line (in this case, $z\approx 0.9$). 
  Probability distributions here and in later figures are normalized
  so that ${\rm max}[P(\log\bar N)] = 1$.
}
\label{fig:lcdm}
\end{figure}

\begin{figure}[t]
\centerline{\psfig{file=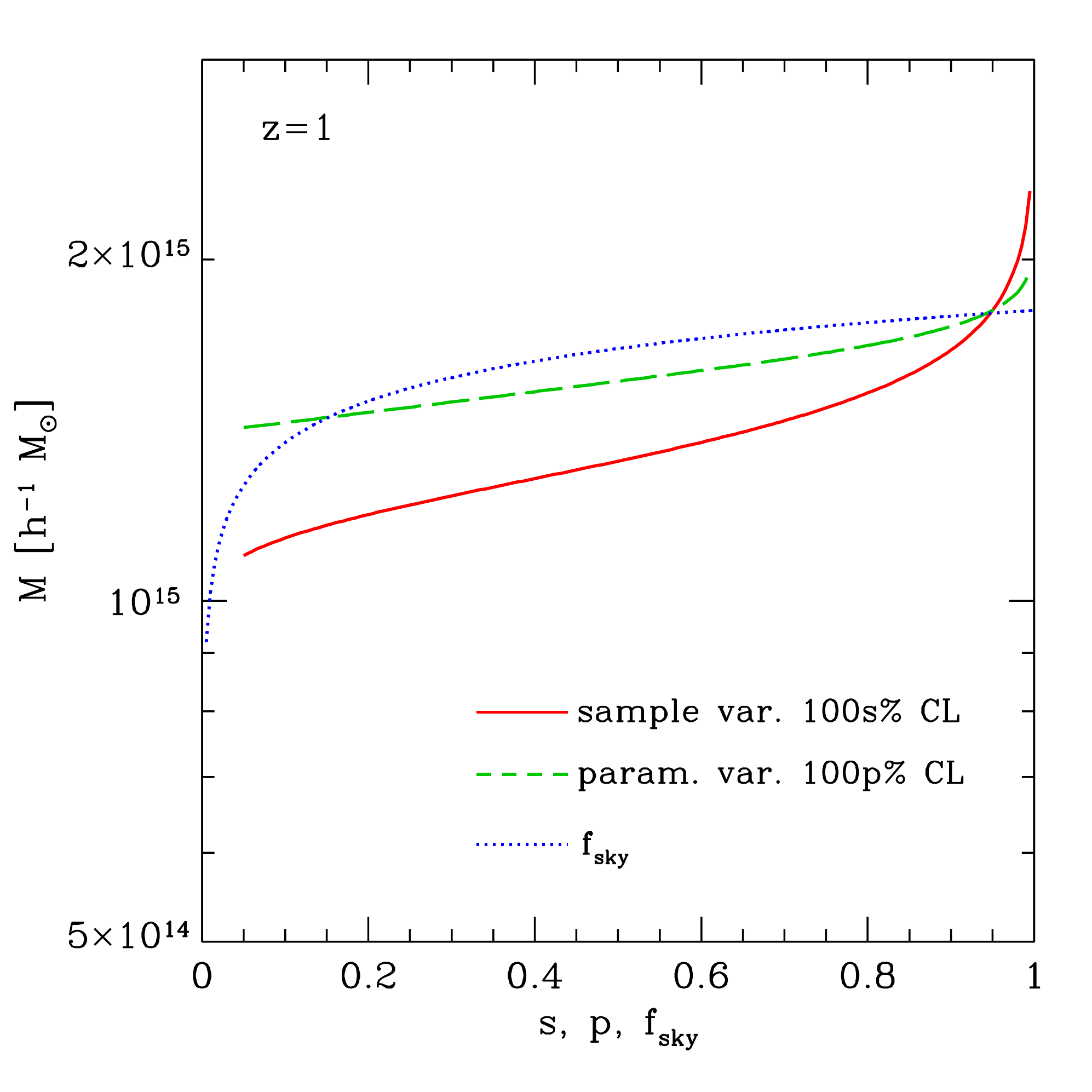, width=3.2in}}
\caption{\footnotesize Dependence of the flat \lcdm\ mass threshold at $z=1$ 
  on sample variance $s$ and
  parameter variance $p$ confidence levels and on the sky fraction $\fsky$.
  Individual variations are computed using the fitting functions of 
  Appendix~\ref{sec:fit} with one parameter varied at a time and the
  remaining parameters fixed to the fiducial values of $s=0.95$, $p=0.95$, and
  $\fsky=1$. 
}
\label{fig:mzscaling}
\end{figure}

\subsection{Model Exclusion}
\label{sec:exclusion}

We begin with the flat \lcdm\ predictions. In Fig.~\ref{fig:lcdm} we show the
posterior distributions of $\log \bar N$ for representative choices of $M$ and
$z$.  As either increases, the mean number drops below unity and the
observation of even a single cluster at that mass and redshift becomes
unlikely.  The dotted vertical line represents the 95\% sample CL
threshold $\bar N_{S.95}(\fsky=1)$.  When $M$ and $z$ are large enough 
that 95\% of the parameter probability
distribution $P(\log \bar N)$ lies below this line, 
we consider the flat \lcdm\ class ruled out at
the 95\% joint CL by an observation of even a single cluster of mass $M$ at
redshift~$z$.

In Appendix~\ref{sec:fit},
we provide a convenient fitting formula for the dependence of these flat
\lcdm\  exclusion masses on redshift, $\fsky$ and the sample and parameter
variance confidence level parameters $s$ and $p$ (see Eq.~\ref{eq:exclusionfit}).  Figure~\ref{fig:mzscaling}
uses these fitting formulas to illustrate how the sample and parameter
variance limits change relative to our default 95\% joint CL full sky limit
with variations in $\fsky$, $s$, and $p$.

\begin{figure}[t]
\centerline{\psfig{file=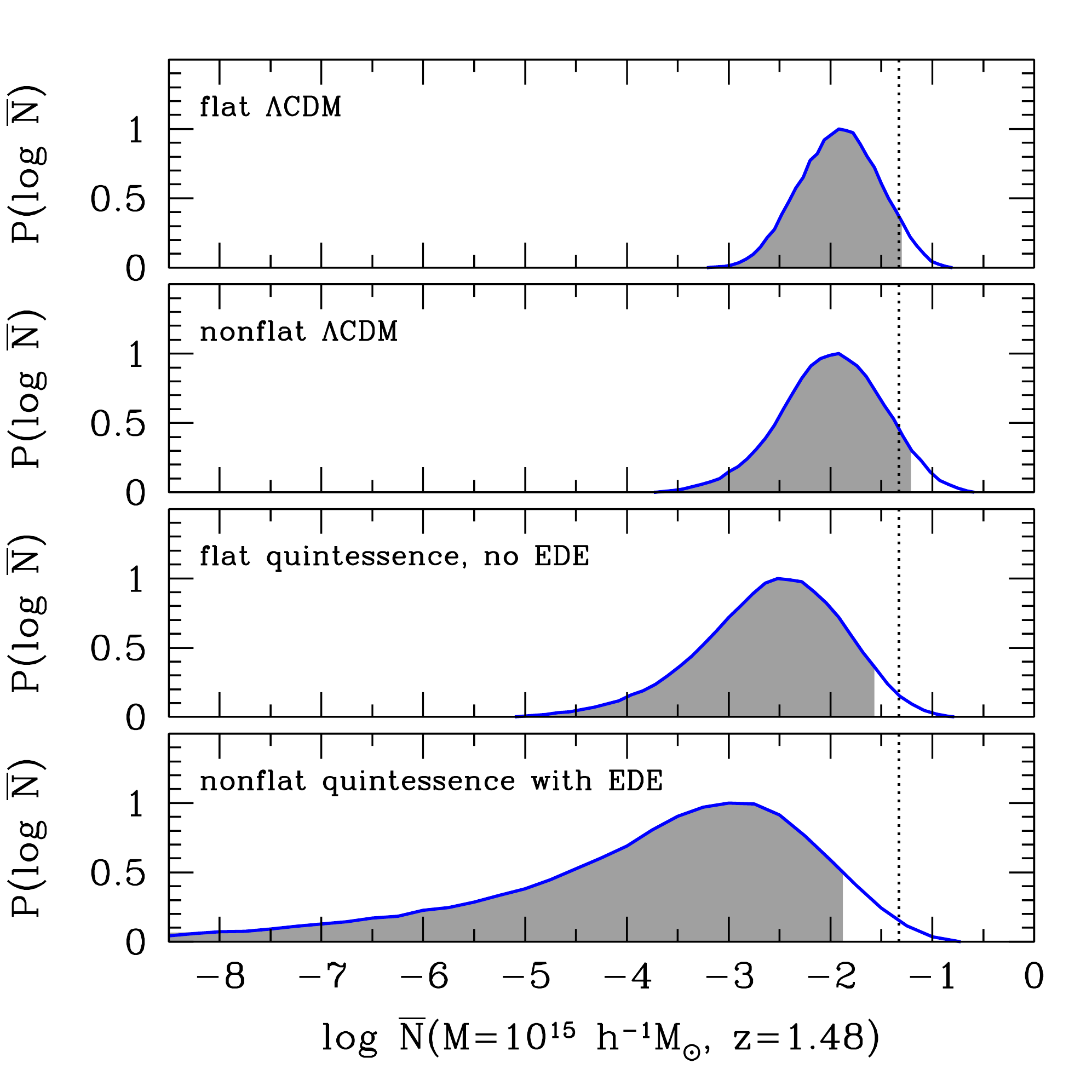, width=3.2in}}
\caption{\footnotesize Predicted mean, full-sky abundance of clusters with
  $M>10^{15}\,\hmsol$ and $z>1.48$, for flat and nonflat \lcdm, flat
  quintessence without early dark energy, and nonflat quintessence with early
  dark energy.  The vertical dotted line marks $\bar N_{S.95}(\fsky=1)$,
  i.e.\  a $10^{15}\,\hmsol$ cluster at $z=1.48$ observed anywhere in the sky
  would exclude all models in $P(\log\bar N)$ to the left of the dotted line 
  with a significance of at least 95\% sample CL. The lowest-$\bar N$
  95\% of each distribution is shaded.
}
\label{fig:model}
\end{figure}

Next we generalize these results to different dark energy model classes.
Figure~\ref{fig:model} shows the predictions for $M=10^{15}\,\hmsol$
and $z=1.48$ which is at the 95\% joint CL for flat \lcdm.  
With nonzero curvature, the confidence level for this choice of $M$ and $z$ 
remains nearly unchanged
at 93\% parameter CL, reflecting the fact that curvature is well constrained in the \lcdm\ context.
  On the other hand, the
parameter confidence level at which models are excluded with 95\% sample CL 
actually increases as the model class widens to flat quintessence without EDE 
($98.8\%$) and to nonflat quintessence 
with EDE ($99.7\%$) (see shading in Fig.~\ref{fig:model}).
 This is in spite of the fact that flat \lcdm\ is included as a special case
 of each of these classes.  In the quintessence classes there are simply
more ways of reducing the growth function at the relevant redshifts through
parameter variations than increasing it.  Hence the low-$\bar N$ tail of the
distribution is highly dependent on the prior placed on the parameters. 

One might therefore worry that the exclusion of quintessence models at the
high-$\bar N$ tail of the distribution might also depend strongly on the prior.  
Parameter choices in the tail of the distribution would then have likelihoods
as good as or better than at the median. This pathology does not occur here.
For example, 
among the 5\% of flat \lcdm\ models in Fig.~\ref{fig:model} with $\bar N>0.051$ 
the best fit model has a likelihood that is worse than the global maximum 
likelihood (for all $\bar N$) by $-2\Delta\ln\mathcal L = 3.3$, consistent with a one-tailed 95\% 
CL.  The best fit quintessence model (nonflat, with EDE) with $\bar N>0.051$ fits the data worse than 
the global maximum likelihood for quintessence by $-2\Delta\ln\mathcal L = 7.5$,
also consistent with the higher confidence for 
exclusion of quintessence models.\footnote{Quintessence models
in the tail of the distribution can actually have a better 
absolute likelihood than \lcdm\ models in the tail or even the median
\lcdm\ model due to the better fit of $w \approx -0.8$
models to the MLCS2k2-analyzed SN data \cite{SDSS_SN}, 
while still being strongly disfavored due
to a large amplitude of structure $A_s$ (or $\sigma_8$).}

For values of
$M$ and $z$ other than those used in Fig.~\ref{fig:model}, the dependence of
$P(\log \bar N)$ on the dark energy model class is similar.  A massive, high $z$
cluster that convincingly falsified \lcdm\ would also falsify {\em all}
quintessence models.
This robustness is
a consequence of the firm upper limit that flat \lcdm\ places on the
quintessence growth function noted in \cite{Mortonson:2009hk} and is
essentially due to the quintessence requirement that $w(z)\ge -1$.
Hereafter we adopt the parameter confidence level of 
flat \lcdm\ for all quintessence cases
to avoid the semantic problem of ruling out quintessence at a higher parameter
confidence than \lcdm\ even though \lcdm\ is a subset of quintessence.

\begin{figure}[t]
\centerline{\psfig{file=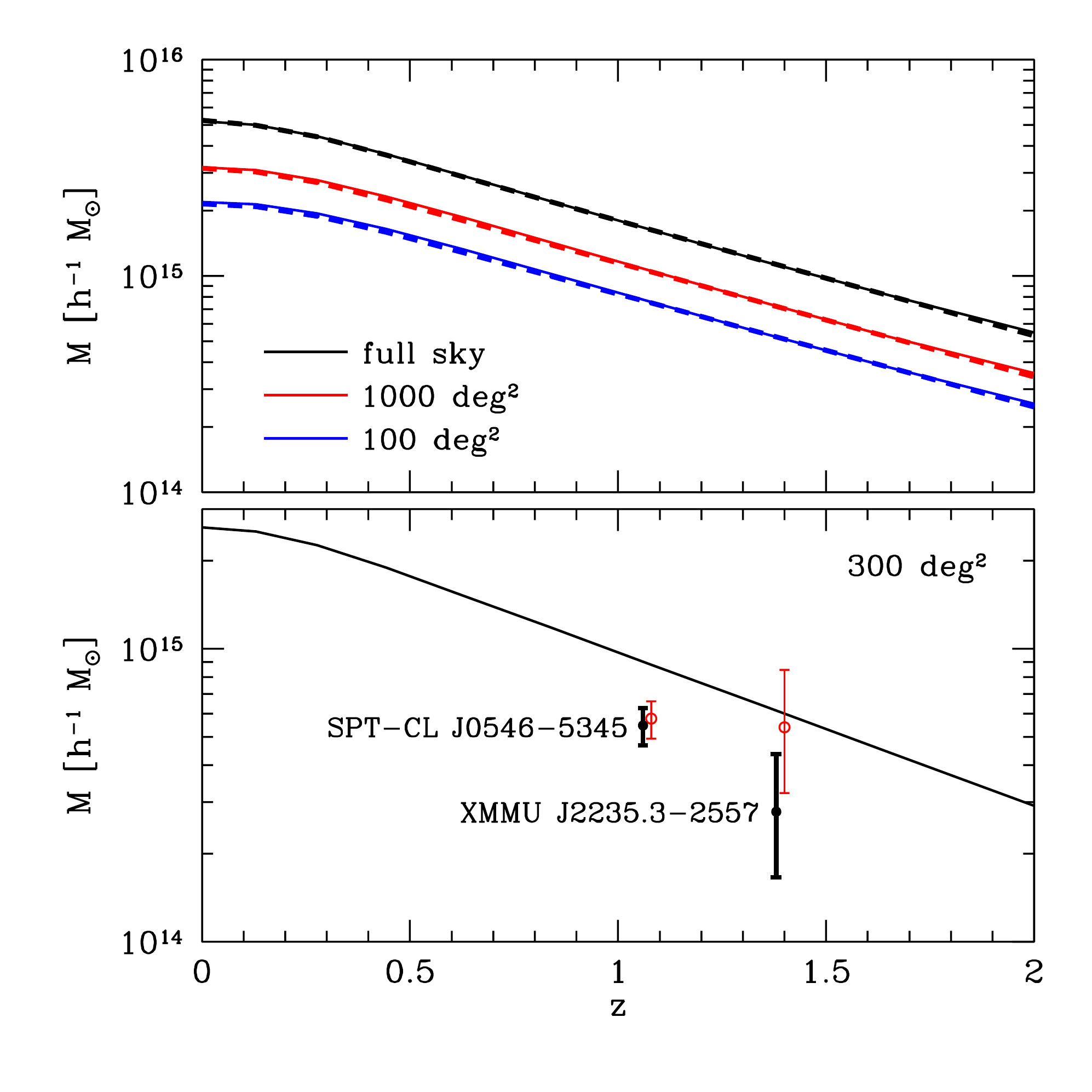, width=3.2in}}
\caption{\footnotesize $M(z)$ exclusion curves.
 Even a single cluster with ($M, z$) lying above the relevant
  curve would rule out both \lcdm\ and quintessence.
 {\em Upper panel:} 
  flat \lcdm\ 95\% joint CL for both sample variance and parameter variance for various choices
  of sky fraction $\fsky$ from the MCMC analysis (thin solid
  curves) and  using the fitting formula from
  Appendix~\ref{sec:fit} (thick dashed curves; accurate to $\lesssim 5\%$ in mass).
  {\em Lower panel:} Two of the most anomalous clusters detected to date, 
  compared with the 95\% joint CL exclusion curve for 300~deg$^2$ which 
  approximates the total survey area for each cluster. We show the 
  X-ray determined masses with and without Eddington bias correction
  (black solid points with thick error bars and red open points with thin error bars, respectively, offset in redshift by $\pm 0.01$ for clarity).
}
\label{fig:mminzmin}
\end{figure}

 In Fig.~\ref{fig:mminzmin} we show the 
95\% joint CL upper limit in the mass-redshift plane for flat
\lcdm.  An observation of one or more clusters at $M$ and $z$ that lie anywhere
above the limit corresponding to a given $\fsky$ would rule out both
\lcdm\ and quintessence.  We further find that the typical realization of the
typical \lcdm\ model, corresponding to the 50\% joint CL, would move the
limiting curve down by a factor of approximately $1.6$ in mass (for
$\fsky=1$).  If we keep the 50\% joint CL and also reduce $\fsky$ to 
correspond to a 300 deg$^{2}$ area, the
mass threshold differs from our fiducial $\fsky=1$ and 95\% joint CL by a
factor of $\sim 3.2$ in mass.  Therefore, to rule out \lcdm\ and quintessence
by our fiducial criteria, the mass of the cluster must be at least $3.2$ times
the typical \lcdm\ prediction for the largest cluster in a 300 deg$^{2}$
survey, and $1.6$ times the prediction for the most massive cluster across the
whole sky.

With these conservative criteria none of the reported high mass, high $z$ clusters
falsify \lcdm\ or quintessence.   The two that provide the most tension with
these model classes are
 SPT-CL J0546-5345 \cite{Andersson:2010vy,Brodwin:2010ig} at  $z=1.07$ 
 which has an X-ray $Y_X$-determined mass of $M_{200} = (8.23\pm 1.21)\times 10^{14}\,\msol$
and XMMU J2235.3-2557 \cite{Stott:2010zr,Mullis:2005hp,Jee:2009nr} 
at $z=1.39$ with an X-ray ($T_X$) mass of 
$7.7^{+4.4}_{-3.1} \times 10^{14}\, \msol$.  These X-ray mass estimates
are consistent with masses obtained by other means such as weak lensing, 
and our most conservative conclusions requiring 95\% joint CL significance 
in the full sky would not be greatly changed by using alternate 
mass proxies.

For a more aggressive interpretation of the data, one can estimate
the effective $\fsky$ values for these measurements.  
They are somewhat subjective
in that the clusters are the most massive ones found in all high $z$
Sunyaev-Zel'dovich (SZ) and X-ray surveys respectively.
The first release of the South Pole Telescope (SPT) SZ cluster survey covered 178 deg$^2$, whereas 
the Atacama Cosmology Telescope SZ survey covered 455 deg$^2$ \cite{Marriage:2010cp} of which 
$\sim 50$~deg$^2$ overlap with the first-release SPT fields.   
On the other hand X-ray surveys have covered some 
  $283~\deg^2$ for $1.0<z<2.2$ \cite{Hoyle}.
  We therefore plot these clusters in Fig.~\ref{fig:mminzmin} (lower panel) 
against an exclusion curve for 95\% joint CL at $300$~deg$^{2}$, 
using $h=0.70$ as assumed in Refs.~\cite{Andersson:2010vy,Stott:2010zr} to 
convert the masses to units of $\hmsol$.\footnote{Specifying $M$ values in
  units of $\msol$ instead of $\hmsol$ has little effect on the widths of the
  $P(\log \bar N)$ distributions even in the quintessence class, suggesting
  that the impact of uncertainties in the Hubble constant due to variations in
  the equation of state near $z\approx 0$ is small \cite{Mortonson:2009qq}.  }
Note that the $M(z)$ level is only
  weakly dependent on $\fsky$ for order unity rescalings
  (see Fig.~\ref{fig:mzscaling}).

   Even under this more aggressive interpretation of
  the exclusion limit, these two clusters do not convincingly rule out \lcdm\ or
  quintessence. 
Although their redshifts and mean masses are somewhat atypical in that 
they exceed the 50\% joint CL exclusion curve, neither cluster is more 
significant than the 95\% joint CL.
 For example, taking the mean reported masses and fixing the 
parameter variance confidence level at 95\%,
  SPT-CL J0546-5345  is only at 44\% sample CL (using the fitting formula of
  Appendix \ref{sec:fit}), i.e.\ it is a typical result for flat \lcdm.  The mean for XMMU J2235.3-2557 
  yields a higher 89\% sample CL, but taking the $1~\sigma$ lower limit on the mass brings the confidence all the way down to 8\%.   Even combining the two using Eq.~(\ref{eq:samp2}) and a joint sky area of 600~deg$^2$ does not improve the confidence.   In fact
  in this conservative test  where thresholds are set to the lowest mass and redshift of the pair, 
  the joint sample confidence level using the mean masses actually decreases to $30\%$.

\subsection{Systematic Shifts}
\label{sec:syst}

Systematic shifts in the observational mass determination, the theoretical mass function,
and SN data analysis techniques can strongly affect the confidence with which
\lcdm\ and quintessence can be excluded. Here we quantify the impact of 
each of these systematic effects on the predicted abundance of high mass, 
high redshift clusters.

Despite numerous recent advances in mass estimation methods, the
determination of cluster masses is still quite uncertain. 
Different methods do not always yield consistent results, and in 
some cases the mass may be systematically over- or underestimated.
Since cluster abundances fall off exponentially with mass at high masses, even
small errors in the estimated masses correspond to large shifts in the
expected number of clusters. 

\begin{figure}[t]
\centerline{\psfig{file=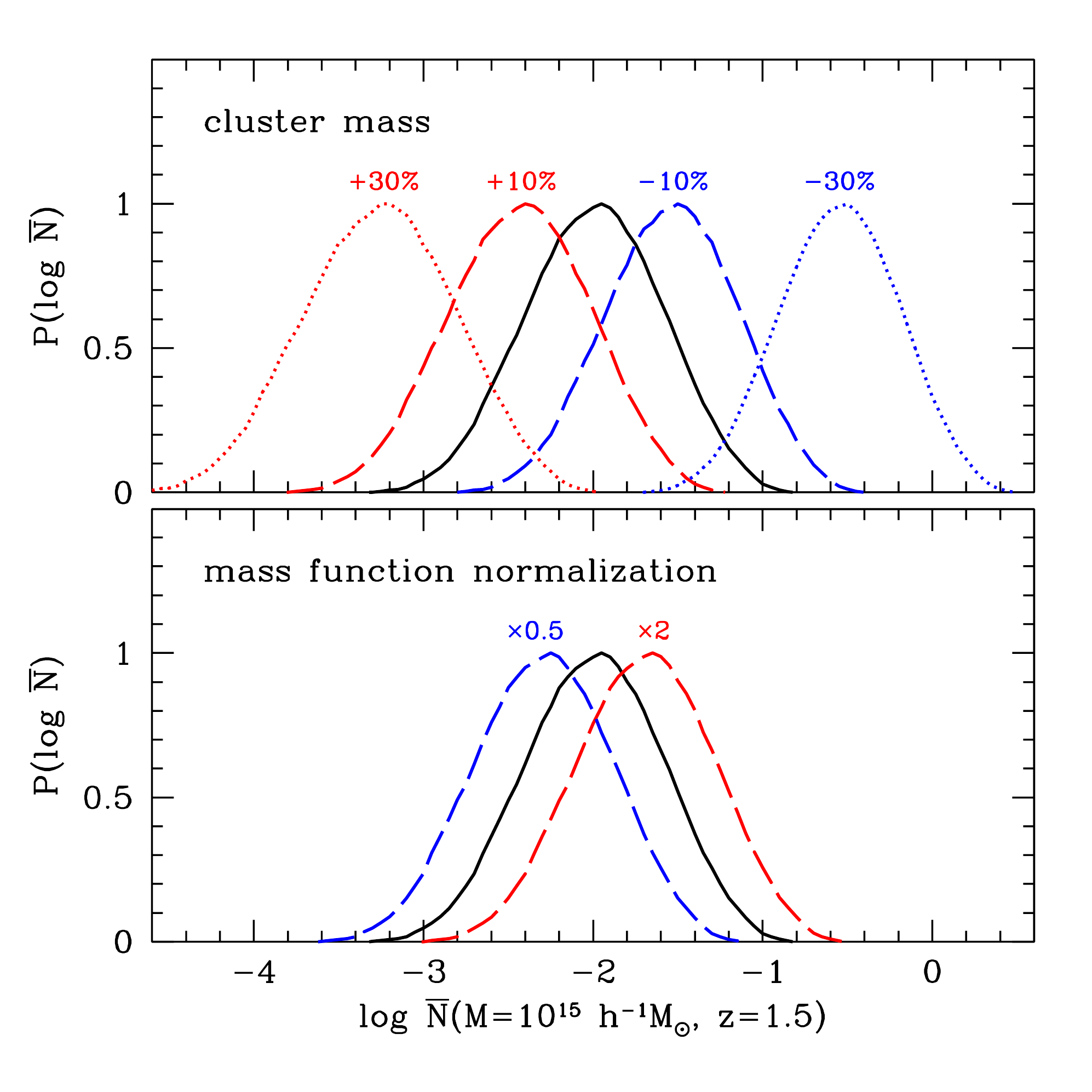, width=3.2in}}
\caption{\footnotesize Impact of systematic errors in cluster mass determination 
and mass function amplitude on the mean number of clusters in the full sky
with $M>10^{15}\,\hmsol$ and $z>1.5$ 
for flat \lcdm.  A fractional change in mass determination can change the 
number of clusters by orders of magnitude.  
Conversely, a factor of two change in the
mass function amplitude near this mass and redshift
changes the mass limits by only a few percent.
}
\label{fig:sys}
\end{figure}

In the upper panel of Fig.~\ref{fig:sys}, we show
the impact on $\bar N$ of changing cluster masses by $\pm 10\%$ or $\pm 30\%$;
these offsets are representative of the range in systematic uncertainty in current
determinations of cluster masses.  Systematic errors in mass are most
important for the rarest clusters due to the increasing steepness of the mass
function.  For $M=10^{14}\,\hmsol$ and $z=0$, a 30\% offset in mass shifts
$\bar N$ by a factor of $\sim 2$, but for the $10^{15}\,\hmsol$, $z=1.5$ case
shown in Fig.~\ref{fig:sys} systematic shifts in mass can change the expected
abundance by orders of magnitude, making an ordinary cluster appear to be
exceedingly unlikely in the context of a given cosmology or vice versa.  In
the $M(z)$ exclusion plane of Fig.~\ref{fig:mminzmin}, these systematic offsets can
be incorporated as simple shifts in the data points.

Estimation of the rarest cluster masses is also subject to Eddington bias,
where selection effects shift the determined masses in a manner that depends on
the cosmology.  If a cluster is selected as anomalous due to the high value of
some observable quantity, e.g.\ X-ray flux and temperature, optical richness, or
SZ decrement, the steep mass function makes scattering from low masses to high
observables more likely than scattering from high masses to 
low observables \cite{Eddington}.  In Appendix 
\ref{sec:eddingtonbias} we discuss two sorts of mass biases associated with
this effect that should not be confused.    For the purposes of comparing to $M(z)$ exclusion curves and
for an observable mass $\Mobs$ that is lognormally distributed
around the true mass, one should correct $\Mobs$ for bias by
\cite{Lima:2005tt,Stanek:2006tu}
\begin{equation}
\Delta \ln M  = {\gamma \over 2} \sigma_{\ln M}^2 \, .
\label{eq:numbermassshift}
\end{equation}
Here $\gamma$ is the local logarithmic slope of the mass function $dn/d\ln M \propto M^\gamma$.
In Appendix~\ref{sec:fit} we provide an approximate expression for 
$\gamma(\bar N,z)$ in 
Eq.~(\ref{eq:alphafit}).  Note that for our default 95\% joint
CL constraint with $\bar N = 0.051$ and $z\sim 1$, $\gamma \approx -8$.  For
$\sigma_{\ln M}=0.3$ this bias is $\Delta \ln M \approx 0.36$ and can have a
substantial impact on the CL level of exclusion should such a high mass cluster
ever be found (see Fig.~\ref{fig:mminzmin}).
The logarithmic slope is much steeper at
this high level of exclusion than the typical expectation for the most massive
cluster in 300 deg$^{2}$ of $\bar N=95$ where $\gamma \approx -5$ at $z \sim
1$.

Specifically, to correct for Eddington bias in placing an observed cluster
whose mass-observable relation implies $\ln M=\ln M_{\rm obs} \pm \sigma_{\ln
  M}$ on the $M(z)$ exclusion plots, one does the following.  Take the implied
$\gamma$ for the $M_{\rm obs}$, redshift, and parameter confidence level $p$
using Eq.~(\ref{eq:Nmp}) and Eq.~(\ref{eq:alphafit}), and evaluate the shift
in mass due to number bias using Eq.~(\ref{eq:numbermassshift}).  Then
take the mean mass and confidence limits and shift them down by this bias
factor.  If the cluster still lies in the excluded region of the $M(z)$ plane,
then it falsifies \lcdm\ and quintessence at the chosen confidence level.
This procedure assumes that the mass function slope is approximately the 
same at $M_{\rm obs}$ as it is for the true cluster mass, which 
holds as long as the scatter $\sigma_{\ln M}$ is not too large.

In Fig.~\ref{fig:mminzmin} we show examples of the Eddington bias correction assuming flat \lcdm\
for SPT-CL J0546-5345 and XMMU J2235.3-2557 where we take the reported mass
errors as a proxy for $\sigma_{\ln M}$.  These examples should only be taken as
illustrative since not all of the sources of mass error are 
lognormally distributed or random. Note that the large mass errors for
the higher redshift cluster and the steeper slope of the mass function both contribute to a bias that is as
large as the statistical errors, although the bias correction from 
Eq.~(\ref{eq:numbermassshift}) may be somewhat 
overestimated given the large $\sigma_{\ln M}$ as noted above.
Taking this bias estimate at face value, the sample variance CL is reduced 
drastically for the corrected mean mass to $<1\%$ for 95\% parameter confidence and $300~\deg^2$, 
whereas the significance of the SPT cluster only falls to $33\%$ given 
its smaller reported mass error.
Relative to \lcdm, quintessence models on average predict that massive, 
high redshift clusters are rarer, resulting in a steeper logarithmic slope 
$\gamma$ and a larger bias correction.

Finally, the mass \emph{definition} used in the theoretical predictions must be
chosen to correspond to a quantity that is tightly correlated with the
observables and consistent with the simulation-calibrated mass function. 
In particular, the
scatter between halo masses in simulations using spherical overdensity and
friends-of-friends halo definitions is large and asymmetric, and the
difference in mass definitions for a single halo can be a factor of two or
more \cite{Tinker:2008ff,Lukic:2008ds}.

Compared with systematic errors in cluster masses, the impact of systematic
errors in the amplitude of the mass function near the relevant mass 
and redshift thresholds
is far less severe.  In particular, although the effect on the mass function of
generalizing to dynamical dark energy models is still largely untested 
by simulations, even
a factor of 2 change in the mass function normalization [i.e.\ $A$ in
  Eq.~(\ref{eq:mf2})] has less impact than a 10\% offset in mass
(Fig.~\ref{fig:sys}, lower panel).  For reference, neglecting the redshift
dependence of the mass function parameters $\{A,a,b\}$ decreases
the amplitude by a smaller $40\%$ shift for \lcdm\ (equivalent to a $3\%$ mass
offset) near $M=10^{15}\,\hmsol$ and $z=1.5$. Following the suggestion of
Ref.~\cite{Tinker:2008ff}, we also test the impact of replacing the redshift
dependence of the mass function parameters with dependence on the growth
function.  Specifically, for a redshift threshold $z$ we evaluate the 
parameters $\{A,a,b\}$ in Eq.~(\ref{eq:mf2}) 
at a different redshift $\tilde z$ satisfying
\begin{equation}
(1+\tilde z)^{-1}G_{\Lambda}(\tilde z)=(1+z)^{-1}G(z)\,,
\end{equation} 
where $(1+z)^{-1}G_{\Lambda}(z)$ is the density 
growth function for a fiducial flat \lcdm\ model.
This modification has a negligible effect on $P(\log \bar N)$ and changes 
the abundance predicted for individual models by $<10\%$ even in the most 
general class of nonflat quintessence models with EDE.

Likewise, extrapolation of the mass function to masses and redshifts outside
the range calibrated to simulations should have a subdominant effect on the
overall systematic errors.  For the mass function we use here, the simulations
of \cite{Tinker:2008ff} probe the range $0.4\lesssim \sigma\lesssim 4$ 
at $z\lesssim 2$ to better than $\sim 5\%$ accuracy. 
For the median \lcdm\ model, the
lower limit of this range corresponds to a maximum mass 
$M\approx 3.1\times 10^{15}\,\hmsol$ at $z=0$, and 
$M \approx 1.0 \times 10^{14}\, \hmsol$ at $z=2$.  
For the 95\% parameter CL \lcdm\ models used to construct
the exclusion curve in Fig.~\ref{fig:mminzmin} these masses are slightly
higher: $M\approx 3.7\times 10^{15}\,\hmsol$ at $z=0$ and $M\approx 1.2 \times
10^{14}\,\hmsol$ at $z=2$.  Thus at high redshift the 95\% joint confidence exclusion curves in
Fig.~\ref{fig:mminzmin} require an extrapolation of up to a factor of $\sim 4$ in
mass for $\fsky=1$ and a factor of $\sim 2$ for 300 deg$^{2}$.  On the other
hand, Hubble volume light cone simulations show no strong deviations from this
mass function \cite{Jenkins:2000bv} from which one can infer that the scaling
holds at least to order unity down to $\bar N \sim 1$; this includes 
the 95\% sample CL rarity for survey areas up to $\sim 2000~\deg^2$.

\begin{figure}[t]
\centerline{\psfig{file=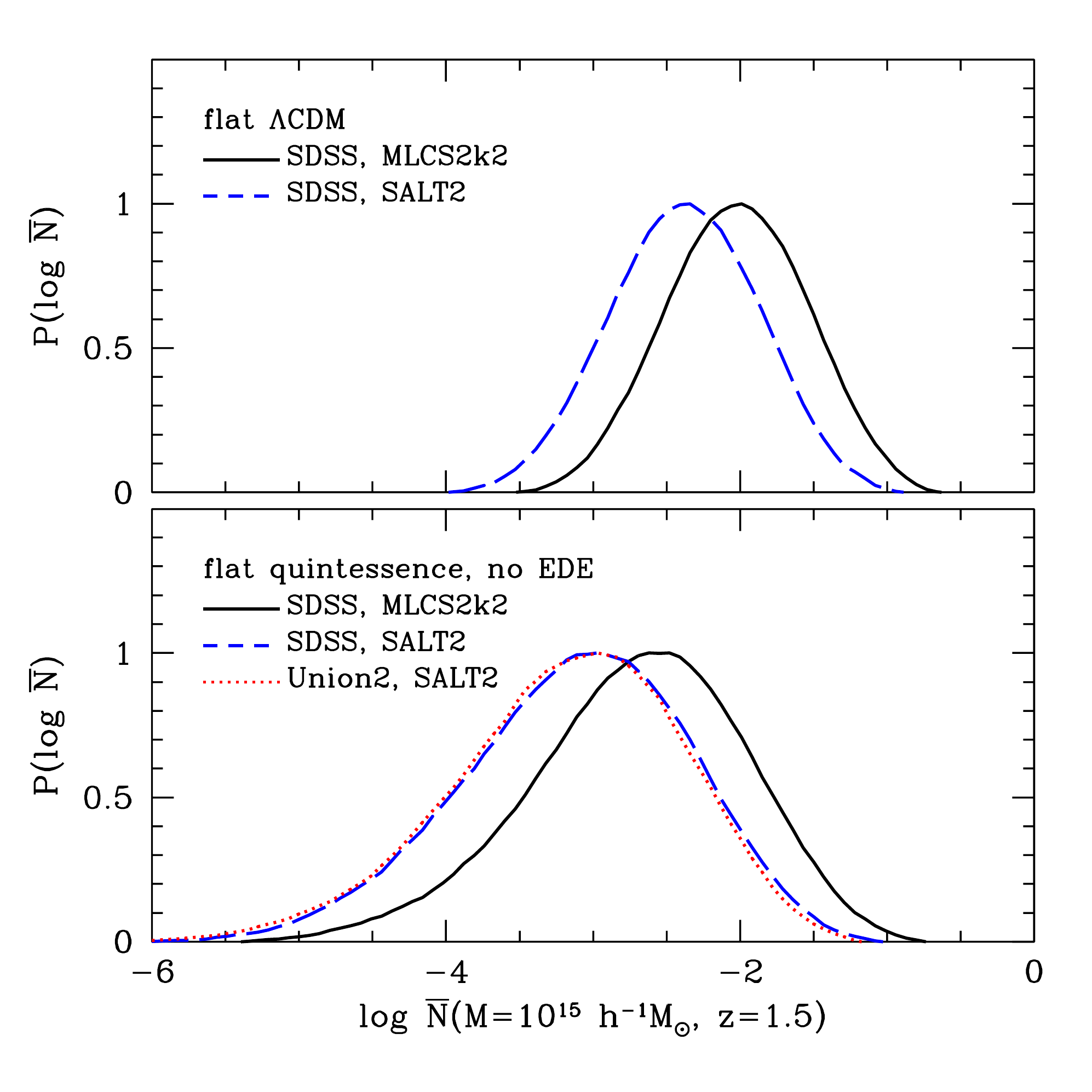, width=3.2in}}
\caption{\footnotesize Effect of SN systematics on $P(\log \bar N)$ at $M=10^{15}\,\hmsol, z=1.5$ for
  flat \lcdm\ (top) and
  flat quintessence without early dark energy (bottom).  The choice of light curve
  fitter when analyzing the SDSS compilation of SN data affects the predicted
  growth history, which leads to a systematic shift in the predicted cluster
  abundance. Switching from MLCS2k2
  to SALT2 has an effect comparable to increasing the mass threshold by $10\%$ 
  (see Fig.~\ref{fig:sys}).
  The bottom panel also shows quintessence predictions 
  using the Union2 compilation of SN data, analyzed with the
  SALT2 method, which are almost identical to the SALT2 predictions with the
  SDSS SN compilation. 
}
\label{fig:sn}
\end{figure}

Systematic errors in the data analysis that propagate into the posterior 
distributions  $P(\log \bar N)$ can also change the confidence at which 
models can be excluded. The largest systematic effects from 
these data sets at present appear to come from the analysis of the SN data; 
in particular, the choice of method for fitting SN light curves (specifically,
MLCS2k2 or SALT2) has been 
shown to affect constraints on a constant dark energy equation of state 
at the level of $\Delta w\sim 0.2$ \cite{SDSS_SN}.  While this specific systematic error will likely be reduced as its 
causes are better understood (e.g.~\cite{Foley:2010mm}), we have adopted
the MLCS2k2 technique for our main results since it provides the more conservative
constraints for assessing exclusion of \lcdm\ and quintessence.

Even for flat \lcdm, the choice of SN methodology affects cluster 
abundance predictions.
  Figure~\ref{fig:sn} (top panel) shows a factor of 2 difference in  
abundance for $M=10^{15}\,\hmsol$ and $z=1.5$. The offset between MLCS2k2 and 
SALT2 varies with the mass and redshift thresholds, but corresponds to 
an approximately constant shift of $10\%$ in the effective mass threshold
for all $z<2$. In Appendix~\ref{sec:fit} we describe how to account for 
this $10\%$ shift in our fitting formulas.
For \lcdm, this difference is mainly due to the preference 
for lower $\om$ when using the SALT2 light curve fitter in place 
of the MLCS2k2 method.  The lower $\om$ also drives down the present day normalization for
fixed initial curvature $A_s$.
The best fit values of $(\om,\sigma_8)$ are 
$(0.29,0.83)$ with MLCS2k2 and $(0.27,0.81)$ with SALT2 (including the 
CMB, BAO, and $H_0$ constraints as well as SN data).
  Using the SALT2 analysis in fact alleviates some tension between the CMB
  and SN data in flat \lcdm. 
  
  The SN distances estimated using the SALT2 method, unlike MLCS2k2, depend on 
an assumed cosmological model and so compiled data sets analyzed assuming 
\lcdm\ formally should not be applied to quintessence \cite{SDSS_SN}.  However
it is both instructive and common practice to do so to approximate  the impact on
dark energy constraints. As in flat \lcdm, using the SALT2 method for flat quintessence
with no EDE on the same
data set lowers the predicted number of clusters (see Fig.~\ref{fig:sn} bottom panel) by a factor of 2 for $M=10^{15}\,\hmsol$ and $z=1.5$, corresponding to 
a $\sim 10\%$ shift in mass. However, for quintessence models this offset 
is due to differences in the preferred dark energy parameters as well as 
in $\om$.
We also show in Fig.~\ref{fig:sn} the impact of switching the SN data to the Union2 compilation, 
which also used the SALT2 method \cite{Amanullah:2010vv}.
Note that the cluster predictions for the two SALT2 cases 
are nearly identical, despite using different sets of SNe.

 We thus have good evidence that flat \lcdm\ exclusion curves with
 either SN light curve fitter provide the same implications for quintessence.  
 The main difference is a shift in the median
 \lcdm\ prediction.  Using the MLCS2k2 light curve fitter predicts higher
 numbers and hence is more conservative for our exclusion analysis.
 Translated into masses, the switch to SALT2 corresponds to a 10\% decrease
 in the mass of the $M(z)$ exclusion limits of
 Fig.~\ref{fig:mminzmin}.  This $10\%$ shift moderately increases the significance of observed clusters.
Using the mean X-ray masses without
Eddington bias correction, and taking the 95\% parameter CL for a $300~\deg^2$ area, 
the sample variance significance shifts from $44\%$ to $62\%$ for SPT-CL J0546-5345 and from $89\%$ to $94\%$ for XMMU J2235.3-2557.

\section{Discussion}
\label{sec:discussion}

In this paper we have analyzed predictions for the abundance of 
massive, distant clusters using an observationally complete basis for
the quintessence paradigm.  Physically, this paradigm assumes that dark energy
is a non-interacting canonical scalar field.  Phenomenologically, quintessence
is a spatially smooth component of energy density compared with dark
matter below the horizon scale with an equation of state $-1\le w(z)\le 1$.

We have shown that any observation that purports to rule out $\Lambda$CDM from
the existence of massive clusters at any redshift also rules out quintessence,
since quintessence models can suppress but not enhance the
abundance of rare clusters compared to \lcdm.  This conclusion still
holds if dark energy is a non-negligible fraction of the total density 
at high redshift. Once normalized to the CMB, quintessence models
can only reduce the number of clusters
(cf.~\cite{Bartelmann:2005fc,Sadeh:2007iz,Francis:2008md,Grossi:2008xh,Francis:2008ka,Pace:2010sn}).
 
We have provided convenient fitting functions that can be used to evaluate the
confidence level of exclusion of a class of dark energy models due to the
observation of a cluster of a given mass at a given redshift.  In doing so,
we have accounted for two sources of variance: {\em parameter variance}, that
current data allow cosmological parameters to take a range of
values, and {\em sample variance}, the Poisson noise in counting rare
objects in a finite volume. Our formulas can also be used to quickly 
evaluate the expected number of clusters in \lcdm.

The single most important element of any claim of model exclusion due to 
observation of a massive, high redshift
cluster is the robustness and accuracy of the mass measurement.  In
particular, it is important to account for Eddington bias, the fact
that the steep mass function will cause lower mass objects to scatter into a
sample defined by thresholds in observable proxies for mass. We include
corrections for Eddington bias in our analysis, 
and clarify the difference between the two types of mass
shifts found in the literature under this name.

When phrased in terms of shifts in the limiting mass, other systematic 
effects are relatively minor in comparison.  For example, 
order unity variations in the mass function amplitude correspond to 
$< 10\%$ changes in the exclusion mass.  Likewise the difference 
between predictions from SN data fit with the SALT2 and MLCS2k2 methods, 
which produces a systematic shift of $\Delta w \sim 0.2$, also 
corresponds to a $10\%$ effect in mass.
 
Finally, we have seen that the interpretation of cluster limits depends
strongly on the effective survey area in which the clusters were selected,
whereas the actual sky area of the data is often much smaller.  The most
conservative interpretation of the most massive cluster in a survey is that
there is at least one such object in the whole sky. Interpreted in this
fashion, none of the clusters reported in the literature can be deemed to
falsify \lcdm\ or quintessence.  Even when interpreted at an  
estimated few hundred square degrees for the effective area, these clusters
fail to convincingly falsify either paradigm.

 Our results differ qualitatively from those in
Refs.~\cite{Jee:2009nr,Holz:2010ck,Cayon,Hoyle} which claim that the observed massive,
high redshift clusters rule out the \lcdm\ paradigm at $\sim 2$--$4~\sigma$. 
The different conclusions can be explained by the fact that these works do not undertake a full treatment of parameter variance, do not correct the observed masses for Eddington bias, and/or use different mass measurements.  Moreover, for single cluster analyses Refs.~\cite{Jee:2009nr,Holz:2010ck,Cayon} assume an effective sky area that, in retrospect, is inappropriately small.
For XMMU J2235.3-2557, a shift in the effective sky area
from 11 deg$^{2}$ to 300 deg$^{2}$ alone accounts for a factor of 1.7 in the exclusion 
mass \cite{Hoyle}.
On the other hand, we do not consider the implications of the full high redshift 
cluster catalog here ({cf}. \cite{Hoyle}).

If in the future a robust case can be made that a massive cluster falsifies
both the \lcdm\ and quintessence classes of models, then at least one
cornerstone of modern cosmology must be incorrect: either 
the initial conditions are
non-Gaussian, dark energy has non-canonical phantom behavior with $w<-1$, dark
energy is not smooth even below the horizon, 
or dark energy interacts with the other components of
the universe.  The latter possibility includes both modified gravity scenarios 
and models where the scalar field responsible for the accelerating
universe interacts with dark matter 
(e.g.\ \cite{Manera:2005ct,Das:2005yj,Abdalla:2007rd,Baldi_Pettorino}).  
Note that
while changing the collisionless cold dark matter aspect of the cosmological
paradigm can also change the cluster abundance, adding a massive neutrino
component can only further suppress the cluster abundance.

Primordial non-Gaussianity, which typically skews the initial distribution of
density fluctuations, can also in principle explain the existence of 
rare, massive, high redshift clusters \cite{Chiu:1997xb}. For example, in the
best-studied local model of primordial non-Gaussianity described by the
parameter $\fnl$ \cite{Komatsu_Spergel}, positive $\fnl$ would increase the
number of clusters relative to $\fnl=0$ (e.g.\ \cite{Sefusatti}).  
However, to substantially 
change the abundance of high $z$ clusters, a large positive
value of $\fnl$ ($\sim 400$) seems to be required
\cite{Jimenez:2009us,Sartoris:2010cr,Cayon,Hoyle}, which, unless one resorts
to postulating more complicated models with scale-dependent non-Gaussianity,
is firmly ruled out by the combination of CMB \cite{wmap7,Smith_fNL} and
large scale structure \cite{Slosar_etal,Afshordi_Tolley,DeBernardis:2010kc}
constraints.

Solutions involving dark energy also run into difficulties if the anomalous
clusters appear only at high redshift.  Typical solutions such as phantom
(i.e.\ $w(z)<-1$ at any redshift) or clustered dark energy 
(e.g.~\cite{Manera:2005ct,Nunes:2005fn,Abramo:2009ne,Creminelli:2009mu})
affect cluster abundances at low redshift as much as or more than at high 
redshift given that the universe has only begun accelerating at
$z\simeq 0.5$.  The same is true of interacting dark energy or modified
gravity scenarios where dark energy effectively mediates an enhanced
attractive gravitational force (e.g.~\cite{Schmidt:2008tn}).
Thus models constructed to explain anomalous high redshift clusters 
while satisfying the CMB and expansion history constraints may still 
be ruled out by the local X-ray cluster 
sample \cite{Ebeling:2000zk,Boehringer:2004fc} or intermediate 
redshift samples (e.g. \cite{Rozo,Vikhlinin:2008cd}).

The standard cosmological paradigm has passed
increasingly stringent tests over the last
two decades.  Current measurements of the expansion history are precise 
enough to make sharp predictions for cosmological structure formation.  
These predictions enable qualitatively new tests
with which the standard paradigm and its extensions can be potentially falsified.
Specifically, the masses of
distant clusters must not be greater than a well-determined number set by
the standard \lcdm\ model if dark energy is a non-interacting canonical 
scalar field with equation of state $w \ge -1$, and if the initial 
conditions are Gaussian.  Thus if increased survey coverage and 
improved cluster mass determination are found to strengthen claims of
clusters that are substantially more massive or more distant than predicted in
\lcdm, then not only specific dark energy model incarnations but the whole  
quintessence paradigm would be falsified.

\vspace{1cm} {\it Acknowledgments:} We thank Tom Crawford, Andrey Kravtsov, and Eduardo Rozo for many
useful discussions.  MJM was supported by CCAPP at Ohio State.  WH was
supported by the KICP under NSF contract PHY-0114422, DOE contract
DE-FG02-90ER-40560 and the Packard Foundation; DH by the DOE OJI grant under
contract DE-FG02-95ER40899, NSF under contract AST-0807564, and NASA under
contract NNX09AC89G. This work was supported in part by an allocation of
computing time from the Ohio Supercomputer Center.  WH and DH acknowledge
generous hospitality from Centro de Ciencias de Benasque ``Pedro Pascual''.

\appendix

\section{Fitting formulas}
\label{sec:fit}

Here we provide a fitting formula to approximate the $M(z)$ exclusion curves
of Fig.~\ref{fig:mminzmin} and their dependence on the sky fraction and
confidence levels for both sample and parameter variance.  As an intermediate
result, we also provide a fit to the median number of clusters expected above a given
mass and redshift for flat \lcdm, $\bar N_{P.50}(M,z)$ as well as its inverse
$M(\bar N_{P.50},z)$.  These fits generalize the expressions provided in
Ref.~\cite{Holz:2010ck}, which approximated $M(\bar N,z)$ for a single
\lcdm\ cosmology across a more limited range in masses and for disjoint
sets of redshifts. We also give an approximate expression for the 
mass function logarithmic slope $\gamma$ which can be used to estimate
corrections for Eddington bias as described in \S~\ref{sec:syst} 
(see also Appendix~\ref{sec:eddingtonbias}).

We begin by fitting an approximate formula for the 
median $\bar N_{P.50}(M,z)$ extracted from the flat \lcdm\ posterior
distributions.  
In order to ensure that the fitting function does not behave unphysically 
beyond the cases tested, we choose a functional form that is motivated
by the mass function in Eqs.~(\ref{eq:mf1}) and~(\ref{eq:mf2}),
\begin{equation}
\bar N_{P.50} \propto e^{-C(M/M_*)^A},
\label{eqn:fitmotivation}
\end{equation}
where $A$, $C$, and $M_*$ are possibly redshift dependent quantities.  This
form follows by assuming that all terms in the mass function vary slowly with
$M$ except for the exponential, $e^{-c/\sigma^2}$, and that the dependence of
$1/\sigma^2$ on $M$ is well approximated by a power law.  Hence we expect the
fit to apply to rare objects such as clusters.

\begin{figure}[t]
\centerline{\psfig{file=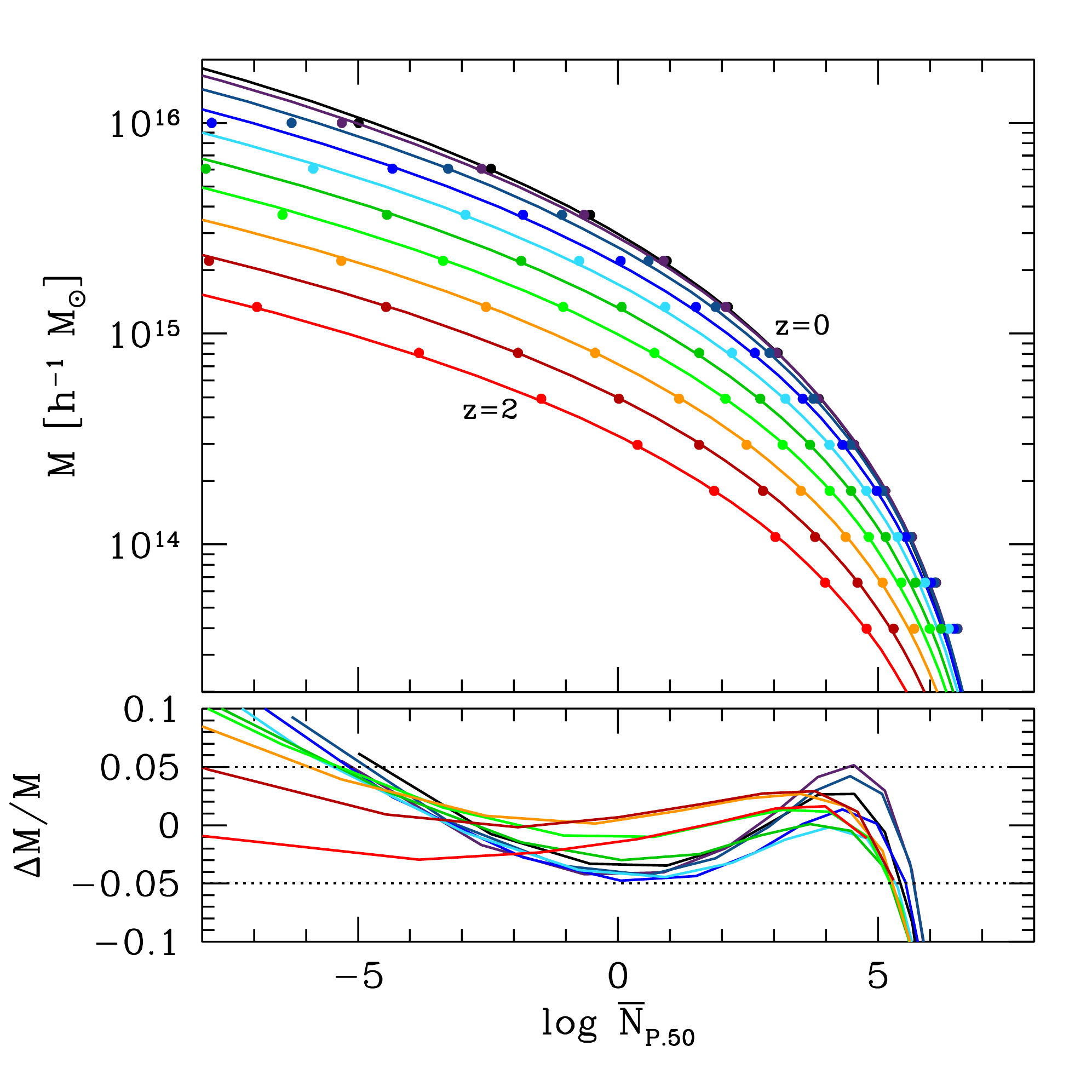, width=3.2in}}
\caption{\footnotesize Accuracy of our fitting formulas for $\bar N_{P.50}$,
  the median \lcdm\ number  of clusters in the whole sky above a given
  mass $M$ and redshift $z$.    {\em Upper panel:} numerical results from the
  MCMC analysis  (points) compared with the approximate
  fitting formula of Eq.~(\ref{eq:fit}) (curves).  Several redshifts are
  shown, with equal spacing in $\ln(1+z)$ over $0\leq z\leq 2$. {\em Lower panel:}
  fitting formula residuals. Dotted lines mark $\pm 5\%$ errors.  }
\label{fig:fit}
\end{figure}

For notational simplicity let us define 
\begin{eqnarray}
m &\equiv &\log[M/(\hmsol)]\,, \nonumber\\
n &\equiv & \log \bar N_{P.50}\,.
\end{eqnarray}
Figure~\ref{fig:fit} shows that the following
expressions are accurate to within $5\%$ in mass over the ranges $14<m<16$,
$0<z<2$, and $-5<n<5$:
\begin{eqnarray}
n(m,z) &=& 7.65 \left[1 - e^{\alpha(z)(m-\beta(z))}\right]\,,\label{eq:fit}\\
&& \alpha(z) = 1.06 - 0.17 e^{-1.3 z}\,, \nonumber \\
&& \beta(z) = 15.565 - 0.1\log\left(7.1 + 10^{5.25z}\right)\,. \nonumber
\end{eqnarray}
In terms of the motivating form of Eq.~(\ref{eqn:fitmotivation}), $C=7.65\ln
10$, $A(z)=\alpha(z)/\ln 10$ and $M_*(z) = 10^{\beta(z)}\,\hmsol$. 
The approximate linearity of $\log M_*(z) \propto \beta(z)$ with $z$ at high redshift
was noted by Ref.~\cite{Holz:2010ck} and indeed equating the scaling of 
$M_*$ with a criterion like $\sigma(M_*)=$ const.\ implies that linearity extends to
$z>2$. This scaling is broken at low redshift mainly because the volume 
saturates and the number above a given $z$ is determined not by the mass function
around $z$ but at a higher effective redshift.

Inverting Eq.~(\ref{eq:fit}) gives the 
cluster mass as a function of redshift and $n$,
\begin{equation}
m(n,z) = \beta(z) + \frac{1}{\alpha(z)}\ln\left(1-\frac{n}{7.65}\right),
\label{eq:fitinv}
\end{equation}
where $\alpha(z)$ and $\beta(z)$ are the same as in Eq.~(\ref{eq:fit}).

The above formulas apply to our predicted cluster abundances using MLCS2k2-fit
SN data in addition to CMB, BAO, and $H_0$ constraints.  The effective $10\%$
shift in mass when using the SN data fit with the SALT2 method instead (see
\S~\ref{sec:syst}) can be simply accounted for by replacing 15.565 with
15.525 in $\beta(z)$ in Eq.~(\ref{eq:fit}):
\begin{equation}
\beta(z) = 15.525 - 0.1\log\left(7.1 + 10^{5.25z}\right) \quad {\rm (SALT2)}.
\end{equation}
 With this change, the
accuracy of the fitting formulas is the same regardless of the method 
used for fitting SN light curves.

The
residuals increase at $n>5$ independent of redshift since we only model
the exponential part of the mass function where halos are rare.
The fitting formulas agree with those of Ref.~\cite{Holz:2010ck} to better than
$\sim 10\%$ in mass for $-2 \le n \le 4$ while avoiding unphysical behavior at $n\le-2$.

\begin{figure}[t]
\centerline{\psfig{file=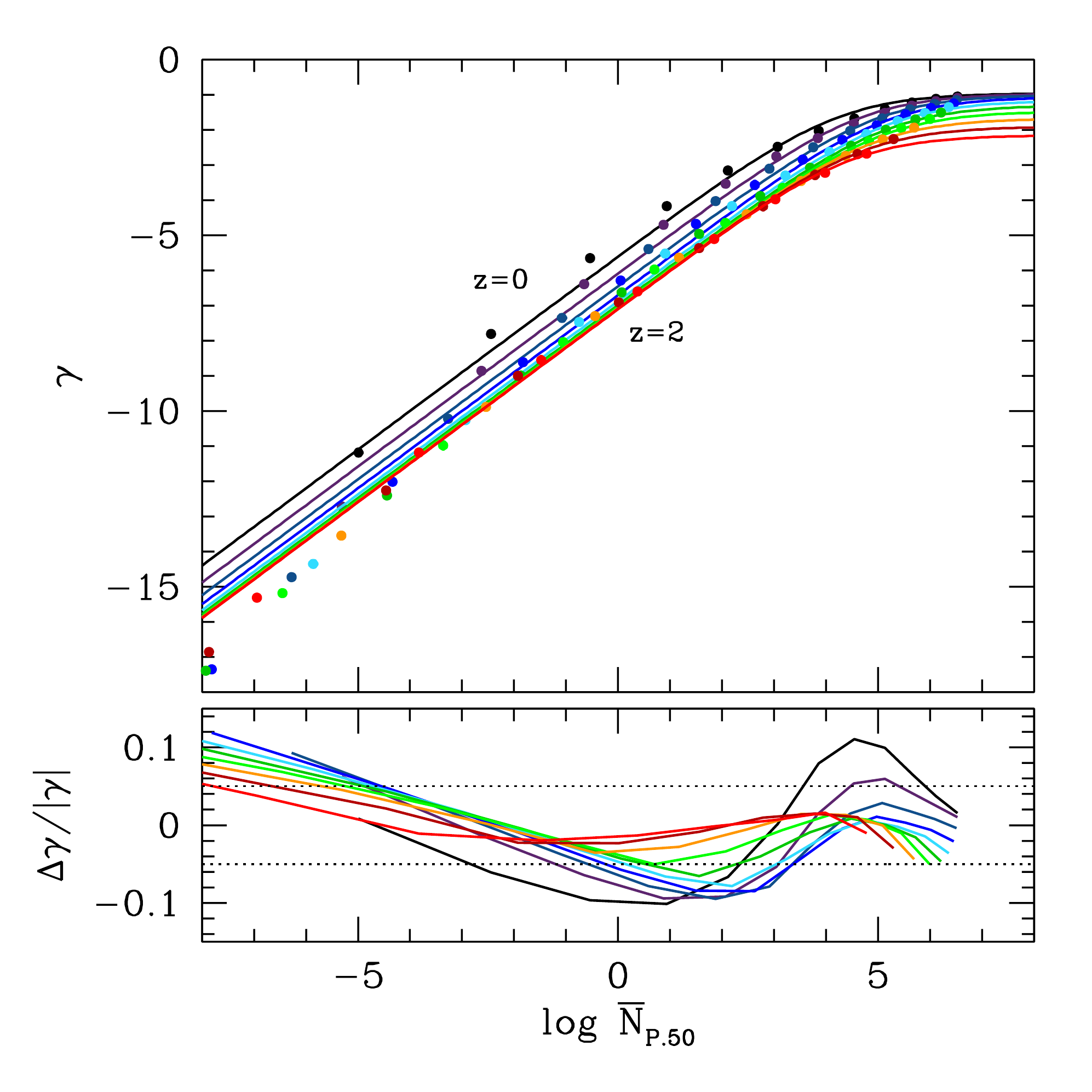, width=3.2in}}
\caption{\footnotesize Fit and residuals for the logarithmic slope of the mass
  function $\gamma$ as a function of median cluster abundance and
  redshift. The fit uses the same points from the numerical calculation 
  as in Fig.~\ref{fig:fit}.  }
\label{fig:alphafit}
\end{figure}

These relations imply that the logarithmic slope of the mass function for
the median \lcdm\ model can be expressed in terms of $\log\bar N$ rather than $M$.
Such a relation has the advantage that the logarithmic slope is a weak
function of redshift at fixed number.  Fitting to numerical results in
Fig.~\ref{fig:alphafit} we obtain
\begin{equation}
\gamma( \bar N,z) = -\ln\left[2.6+1.5 z^2+e^{7.1-1.5\exp(-3z)-1.1\log \bar N}\right]. \label{eq:alphafit}
\end{equation}
Given the relationship between the median $\log \bar N=n$ and $m$
 in
Eq.~(\ref{eq:fit}), this expression gives the Eddington bias correction
as a function of mass and redshift assuming the median \lcdm\ predicted number.
The fit to $\gamma(\bar N,z)$ is equally valid for predictions using 
both the MLCS2k2 and SALT2 SN analyses, although the relation between 
$\log \bar N$ and $M$ differs as described above.

\begin{figure}[t]
\centerline{\psfig{file=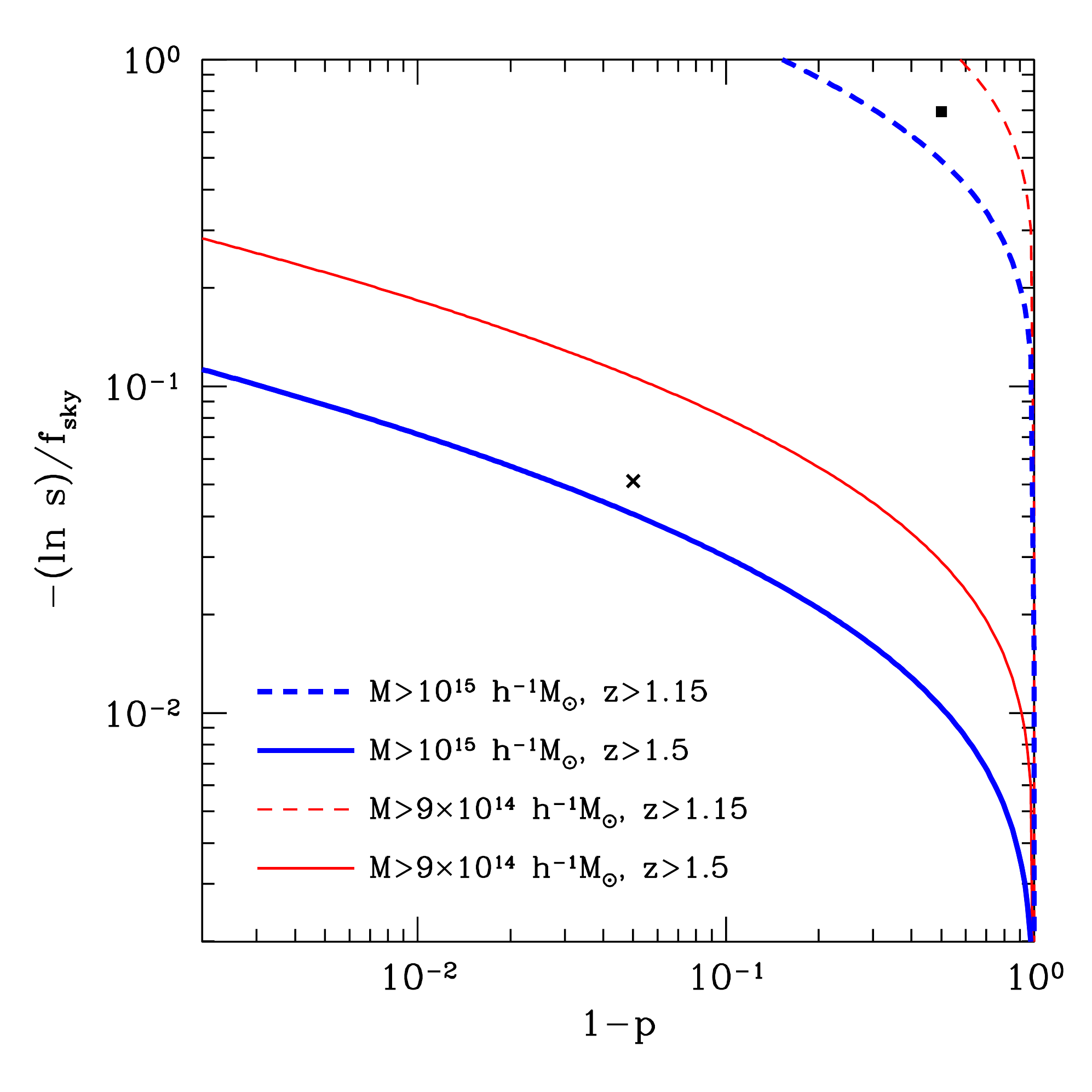, width=3.2in}}
\caption{\footnotesize Combinations of $100s\%$ CL and $100p\%$ CL limits for sample
  variance and parameter variance, respectively, given various mass and
  redshift thresholds.  In the limit of high significance for sample 
  variance ($1-s\ll 1$), 
  $-\fsky^{-1}\ln s \approx \fsky^{-1}(1-s)$. The curves are computed using the
  fitting formulas of Eqs.~(\ref{eq:fit}) and~(\ref{eq:logn}).  The 95\% joint
  CL values used in \S~\ref{sec:predict} and the 50\% joint CL values for 
  $\fsky=1$ are marked with a cross and a square, respectively.
}
\label{fig:sp}
\end{figure}

Changing the parameter variance exclusion level $p$ changes the
relationship between $\bar N$ and $m$ (see Fig.~\ref{fig:lcdm}).  For a higher
confidence level than the median $p=0.5$ the predicted number at a 
fixed mass increases.
Thus in order to find the mass $m$ as a function of $\log \bar N$ for either
exclusion curves or evaluating the logarithmic slope of the mass function, we
need to shift the effective number density at which we evaluate
Eqs.~(\ref{eq:fitinv}) and~(\ref{eq:alphafit}).  
We approximate $P(\log \bar N)$ as a lognormal
distribution for the flat \lcdm\ model class (see Fig.~\ref{fig:lcdm}) with
mean $n$ and width
\begin{equation} 
\sigma_{\log \bar N} = 0.29 - 0.035 n\,,
\end{equation}
independent of redshift (fit to the distributions in Fig.~\ref{fig:lcdm} over 
$-8<n<6$).
Then the relation between $\bar N$ and $m$ and $z$ at some parameter confidence
$p$ is given by
\begin{eqnarray}
\log \bar N_{Pp}(m,z) &=& [{1-0.035\sqrt{2}\,{\rm erf}^{-1}(2p-1)} ]n(m,z) 
\nonumber\\
&& + 0.29\sqrt{2}\,{\rm erf}^{-1}(2p-1).
\label{eq:Nmp}
\end{eqnarray}

The inverse relationship between $n$ and $m$ of Eq.~(\ref{eq:fitinv}),
\begin{equation}
\log \frac{M(z;s,p,\fsky)}{\hmsol} = m(n(s,p,\fsky),z) \, ,
\label{eq:exclusionfit}
\end{equation}
can then be evaluated at
\begin{equation}
n = \frac{\log(-\fsky^{-1}\ln s)-0.29\sqrt{2}\,{\rm erf}^{-1}(2p-1)}
{1-0.035\sqrt{2}\,{\rm erf}^{-1}(2p-1)} \, ,
\label{eq:logn}
\end{equation}
which comes from Eq.~(\ref{eq:Nmp}) using the criterion that the 100$s$\% CL
sets $\bar N_{Pp} = \bar N_{Ss} = -\fsky^{-1}\ln s$.
Equation~(\ref{eq:exclusionfit}) therefore gives the desired exclusion curves
at a given sample variance CL, parameter variance CL, and sky fraction.
The Eddington bias at a given parameter variance confidence, mass, and redshift
can also be evaluated by using Eqs.~(\ref{eq:Nmp}) and~(\ref{eq:fit}) 
in Eq.~(\ref{eq:alphafit}).

These formulas also provide a convenient way to estimate the significance of an
observed cluster: given the cluster mass and redshift, Eq.~(\ref{eq:fit})
approximates the median expected cluster abundance in the full sky, and
Eq.~(\ref{eq:logn}) then determines the corresponding combinations of sample
variance and parameter variance confidence limits for flat
\lcdm. 
Note in particular that $-\fsky^{-1}\ln [s(m,z;p)]$ can be extracted as  a closed form expression.
Figure~\ref{fig:sp} shows examples of these $s$--$p$ relations for
clusters with $M\approx 10^{15}\,\hmsol$ at redshifts $z=1.15$ and $1.5$, 
for which the significance of excluding flat \lcdm\ is near the 50\% joint CL
and 95\% joint CL, respectively. For high significance clusters, 
a $10\%$ change in mass shifts the exclusion curves by a factor of a few in 
$(1-s)/\fsky$ and over an order of magnitude in $(1-p)$, highlighting 
again the importance of accurate mass determination.

\begin{figure}[t]
\centerline{\psfig{file=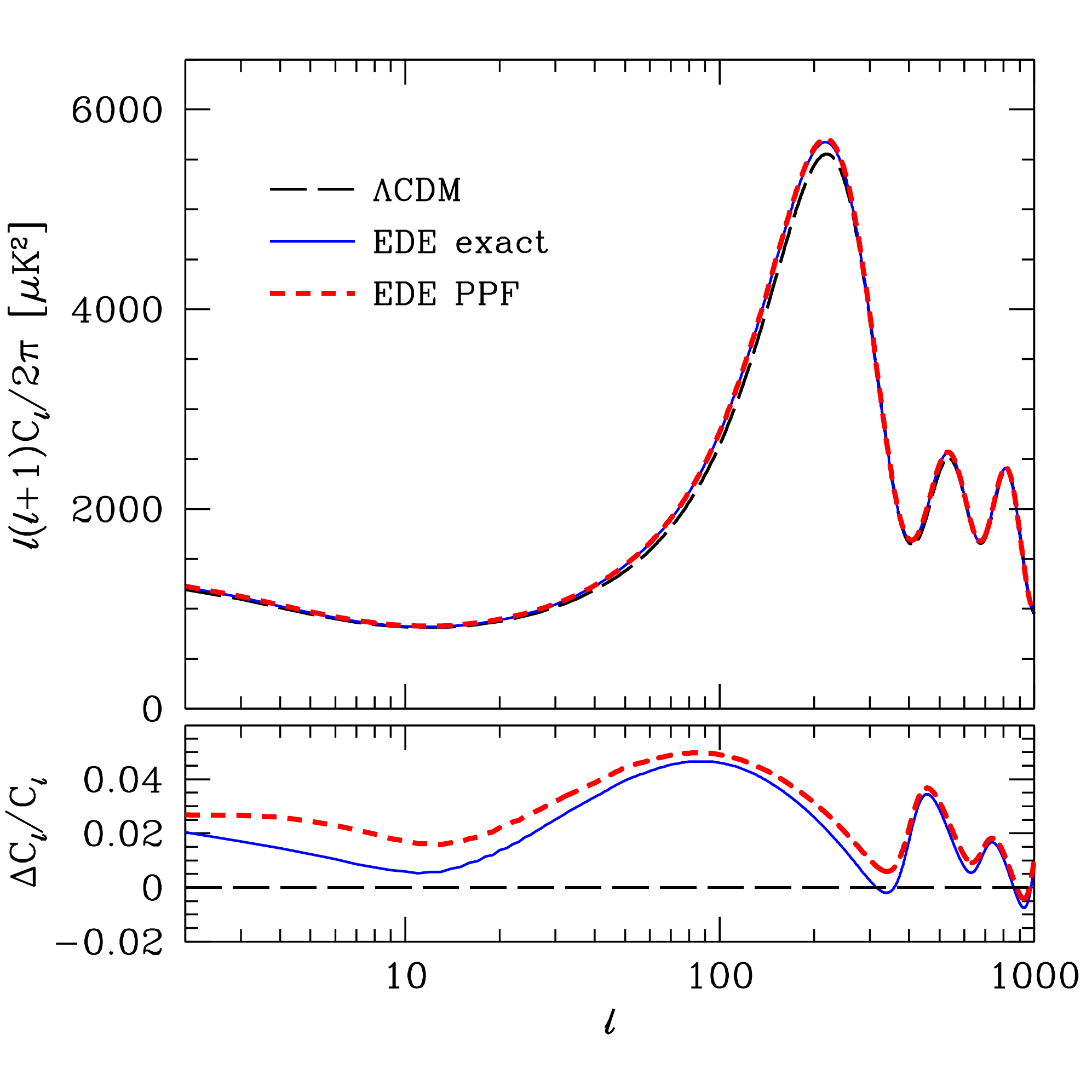, width=3.2in}}
\caption{\footnotesize CMB temperature power spectrum of an EDE model calculated
exactly from the scalar field equations and in the PPF approximation, compared
with a \lcdm\ model with the same matter density, baryon density, distance
to last scattering, and initial curvature power spectrum.  
The transition from clustered to smooth EDE boosts the CMB power spectrum 
around the first peak, and thus matching the WMAP7 normalization 
requires small changes in $A_s$ and $n_s$ from their \lcdm\ values.
}
\label{fig:edetest}
\end{figure}

\section{Normalization and Early Dark Energy}
\label{sec:ede}

High redshift cluster abundance constraints are often phrased relative to the local cluster
abundance by fixing $\sigma_8$ and $\Omega_{\rm m}$. 
Despite how well these two parameters are  constrained in flat \lcdm\ by 
the WMAP7 data, this is not equivalent to normalizing to CMB data.

In particular, the assumption that the value of $\sigma_8$ is well known requires
extrapolating from the measurement at $z_*=1090$ to the present using a
particular growth function, and in general changing the dark energy model
results in a different growth function implying a different value of
$\sigma_8$ for fixed CMB amplitude.  Additionally, the presence of EDE
directly affects CMB fluctuations on the horizon scale at recombination due to
its transition from an adiabatically clustered to relatively smooth component.

Our normalization parameter is $A_s$, the amplitude of the {\em initial}
curvature power spectrum at $k=0.05$\,Mpc$^{-1}$, and we use the CAMB PPF
module to propagate its effects jointly with the effects of EDE on the observable CMB
power spectra.  The PPF approximation retains the quintessence property that
the sound speed of dark energy is equal to the speed of light but implements
the transition from a clustered to smooth component in an approximate manner
\cite{PPF}.

In Fig.~\ref{fig:edetest} we compare
a \lcdm\ model and a quintessence model with EDE which have
the same $A_s$, $n_s$, 
$\Omega_{\rm c} h^2$,   $\Omega_{\rm b} h^2$ and $D(z_*)$.
Specifically, we choose an offset exponential for the quintessence potential
\begin{equation}
V(\phi) = V_0 + A \exp(-\phi/\phi_*)\,.
\end{equation}
For this potential, quintessence  behaves as a tracking field at early times 
with $w\approx 0$ during matter domination, and as a cosmological constant
at late times.  
We first calculate the exact CMB power spectrum for this model and then
compare it with the PPF approximation.
The EDE to matter ratio at $z_*$ is determined by $\phi_*$
 (e.g.~\cite{Ferreira:1997hj}) and
we have chosen it to correspond to 3.3\% while the ratio of
EDE to the total density is 2.4\%.  
The potential offset $V_0$ is set by demanding that $D(z_*)$
remain fixed in a flat universe, and $A$ is set by requiring that 
the field be in the tracking regime well before recombination.

 Note that to zeroth order the CMB power at the $\ell\sim200$
first peak remains largely fixed given the same initial curvature power spectrum, but the small EDE fraction causes
a first order correction.  The decay of gravitational
potentials due to the EDE becoming smooth under the horizon scale leads to 
a small boost in the amplitude.  Conversely, at fixed WMAP7 normalization, the
best fit $A_s$ is slightly reduced, while the best fit $n_s$ increases
to preserve the amplitude of the first peak where 
the EDE is most effective ($k\approx$ 0.02\,Mpc$^{-1}$ compared
with our pivot scale of 0.05\,Mpc$^{-1}$).
Moreover, the boost widens the first peak and since this effect is not
degenerate with changes in $A_s$ and $n_s$ or other cosmological parameters, 
it provides limits on EDE from the CMB alone \cite{Doran:2006kp}.
The PPF approximation captures
these effects to sufficient accuracy for models that satisfy these observational bounds.

\begin{figure}[t]
\centerline{\psfig{file=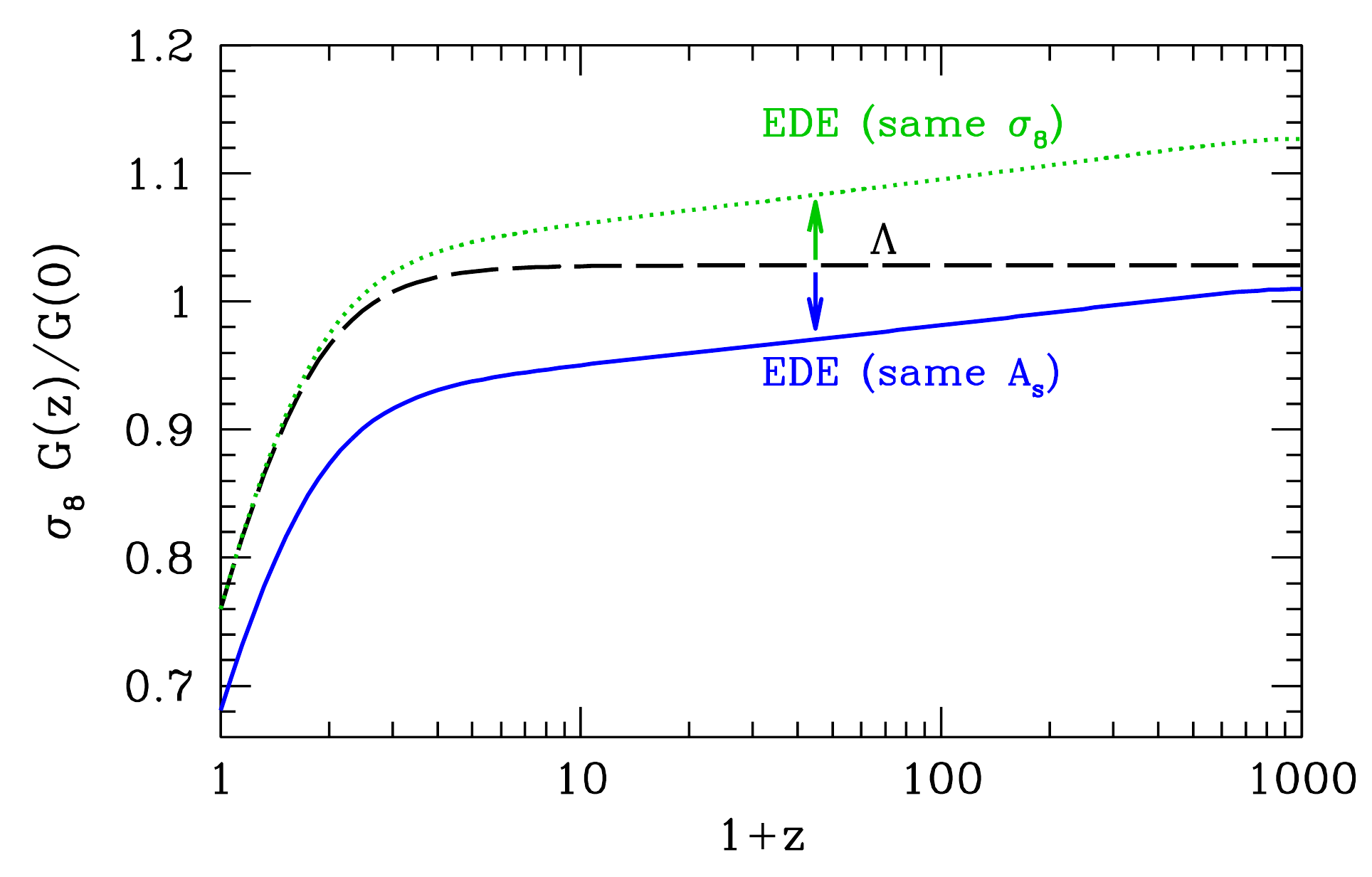, width=3.2in}}
\caption{\footnotesize Illustration of the varying effects of early dark 
energy depending on whether the growth history is normalized at $z=0$ 
by fixing $\sigma_8$ or at $z\sim 10^3$ by fixing $A_s$. 
The \lcdm\ and EDE example models from Fig.~\ref{fig:edetest} are shown
by the dashed black and solid blue curves, respectively; 
the dotted green curve shows the same EDE model with $\sigma_8$ rescaled 
to match its value in the \lcdm\ model.
}
\label{fig:sigma8}
\end{figure}

We obtain $\sigma_8$ and more generally $\sigma(M,z=0)$ for each cosmological model using the PPF 
version of CAMB.  Even the small EDE fraction of our example can have a sizable effect on
$\sigma(M,z=0)$ due to the long lever arm between $z_*$ and $z=0$ over which the
change in the sub-horizon growth rate can act.   
To obtain $\sigma(M,z)$ at $z>0$, we compute $G(z)$ by integrating the differential 
equation for the linear growth function as described in \cite{Mortonson:2008qy}.
Since we use the growth function only to scale backwards from the present epoch,
all high redshift modifications to the transfer function and 
the CMB normalization of $\sigma(M,0)$ through $A_s$ are accounted for, 
including the impact of the EDE clustering transition.  
This procedure is illustrated in Fig.~\ref{fig:sigma8}
(lowest curve).  We then assume that $\sigma(M,z)$ determines the
halo mass function with no explicit dependence on EDE.  
Note that this differs from \cite{Bartelmann:2005fc,Sadeh:2007iz} who
took a spherical collapse motivated mass function with a collapse threshold
$\delta_c$ that depended on EDE.  The modified threshold ansatz was later shown
to be inconsistent with both numerical and analytic results~\cite{Francis:2008md,Francis:2008ka,Grossi:2008xh,Pace:2010sn}.

The alternate approach of assuming a fixed value of $\sigma_8$ leads to 
very different conclusions about the effects of EDE. In that case, the 
similar shapes of the \lcdm\ and EDE growth functions at low redshifts 
imply that the effects of EDE on cluster abundances are small 
(upper curves in Fig.~\ref{fig:sigma8} at low $z$) \cite{Francis:2008md,Francis:2008ka,Grossi:2008xh,Pace:2010sn}. 
However, Fig.~\ref{fig:sigma8} combined with Fig.~\ref{fig:edetest} shows that a 
model with a substantial amount of EDE and $\sigma_8$ fixed to the 
best fit \lcdm\ value would change the 
high redshift normalization $A_s$ and hence the CMB power spectrum normalization.
In this example, the shift is more than 10\% in amplitude, 
whereas the CMB normalization for a given $A_s$ 
is determined to an accuracy of $\sim 1.4\%$
corresponding to the uncertainty in the reionization optical depth.
Conversely, by requiring consistency with the CMB, the effect of EDE on 
{\em reducing} the cluster abundance can be quite large as shown in 
Fig.~\ref{fig:model} (see also \cite{Alam:2010tt,Basilakos:2010fb}).

\section{Eddington Bias}
\label{sec:eddingtonbias}

For a steep mass function, an observable proxy for cluster mass $\Mobs$ 
is biased high compared with the true mass $M$ since it is more likely that 
one of the numerous low mass objects scatters to higher $\Mobs$ than it is 
that a rare high mass object scatters to lower $\Mobs$.  
There are two types of mass bias 
associated with this effect and we clarify their use here.

Consider first the type relevant to the exclusion analysis of the main 
part of the paper.
We seek to find the probability of a given model producing a cluster with observed mass
greater than $\Mobs$ at the given redshift or above.  
The generalization of  $\bar N$, the number of clusters above a given true mass and
redshift, is
\begin{eqnarray}
\bar N_{\rm obs}(\Mobs,z)  
&=&  \int_{z}^{\infty} dz' \frac{4\pi D^2(z')}{H(z')}
\int_{\Mobs}^{\infty} \frac{d\Mobs'}{\Mobs'} \int_{0}^{\infty} \frac{dM'}{M'}
\nonumber\\&&\times
 \frac{dn}{d\ln M}(M',z')P(\ln \Mobs'|\ln M')\,,
\label{eq:nobs}
\end{eqnarray}
where $P(\ln \Mobs | \ln M)$ is the probability density of obtaining an observed mass
$\Mobs$ given a true mass $M$.  
In order to set the probability of finding a cluster of observed mass $>\Mobs$
equal to finding a cluster of true mass $>M$ we require
\begin{equation}
\bar N_{\rm obs}(\Mobs,z)= \bar N(M_N,z)\,.
\end{equation}
The difference between $M_N$ and $\Mobs$ is the {\it number bias} mass shift. 
By setting the probabilities equal, we therefore have the same exclusion confidence
as if we had measured a cluster with true mass $M_N$.

For the case of a lognormal mass observable relation with rms $\sigma_{\ln M}$ that
is small compared with the scale over which the local slope of the mass function
changes, $dn/d\ln M\propto M^\gamma$  and \cite{Lima:2005tt,Stanek:2006tu}
\begin{equation}
\ln M_N = \ln \Mobs + {1\over 2}\gamma \sigma_{\ln M}^2 \,.
\label{eq:numberbias}
\end{equation}
This is the appropriate mass to plot on an $M(z)$ exclusion plot.
Note that $\gamma<0$ so $M_N<M_{\rm obs}$.

There is a second sense of a bias in mass that is commonly used in the
literature.   To place confidence limits on the mass assuming a mass function and
an observed mass $\Mobs$, one can use Bayes' theorem  \cite{Hogg:1998mu,Vanderlinde:2010eb}
\begin{equation}
P(\ln M | \ln \Mobs) \propto P(\ln M) P(\ln \Mobs | \ln M)
\end{equation}
and take $P(\ln M) \propto dn/d\ln M$.   For the same lognormal and constant slope 
assumptions, the posterior mass distribution is a lognormal of the same width
$\sigma_{\ln M}$ and shifted mean 
\begin{equation}
\ln M_M =  \ln \Mobs + \gamma \sigma_{\ln M}^2\,,
\label{eq:massbias}
\end{equation}
which is twice the mass shift required to hold probabilities fixed.  Note that
this {\em mass bias} mass shift is the answer to a statistically different
question.  Here one assumes that the observation is fixed and the mass
function is {\it a priori} correct. One does not account for the probability
of drawing such an $\Mobs$ (or greater) from the mass function and the mass
observable relation.  In other words, Eq.~(\ref{eq:massbias}) is the
appropriate correction for quoting confidence levels for the mass assuming
\lcdm, and Eq.~(\ref{eq:numberbias}) is the appropriate correction for quoting
confidence levels for how the existence of a given cluster might exclude
\lcdm.  It is the latter that we are interested in here.

\vfill

\bibliography{pccluster}

\end{document}